\documentclass[
 amsmath,amssymb,
 aps,
]{revtex4}

\usepackage{braket}
\usepackage{graphicx}
\usepackage{dcolumn}
\usepackage{bm}
\usepackage{color,float}
\usepackage{type1cm}
\usepackage{comment}
\newcommand{\simgt}{\lower.5ex\hbox{$\; \buildrel > \over \sim \;$}}
\newcommand{\simlt}{\lower.5ex\hbox{$\; \buildrel < \over \sim \;$}}

\begin{document}

\author{Daisuke Miki}
\author{Akira Matsumura}
\author{Kazuhiro Yamamoto}

\affiliation{
Department of Physics, Kyushu University, 744 Motooka, Nishi-ku, Fukuoka, Japan
}

\date{\today}

\title{
Non-Gaussian entanglement in gravitating masses:
the role of cumulants
}

\begin{abstract}
We develop an entanglement criterion with third- and fourth-order
cumulants to detect the entanglement of non-Gaussian states.
The efficiency of the entanglement criterion is investigated for gravitating mirrors in optomechanical systems.
We show that the entangled regime of the mirrors is enlarged by the third- and fourth-order cumulants.
We also discuss the limitations of the entanglement criterion for mirrors in a highly non-Gaussian state.
\end{abstract}
\maketitle

\section{\label{sec:level1}INTRODUCTION}

Various efforts have been made to unify gravity and quantum mechanics.
However, an established theory of quantum gravity has not been developed yet.
One of the difficulties in the unification is that there are
few experiments to test the quantum behavior of gravity.
The description of gravity according to the principles of quantum mechanics is also under debate \cite{Penrose, Diosi}.
Feynman \cite{Feynman} discussed the quantumness of gravity from the viewpoint of quantum superposition. 
Based on the recent progress in quantum science, the possibility of verifying whether gravity obeys the superposition principle has been discussed \cite{Aspelmeyer,Schmole,Matsumoto,Lopez,Carney}.
The Bose--Marletto--Vedral (BMV) experiment focusing on the quantum entanglement generated by Newtonian gravity \cite{Bose,Marletto} has
stimulated recent studies (e.g., \cite{Marshman,Belenchia,Christodoulou,Anastopoulos,Grossardt,Kamp,Krisnanda,Nguyen,Miki, Matsumura2}) and discussions \cite{Anastopoulos2,Hall}.

Quantum entanglement is a nonlocal correlation known in quantum mechanics that plays a central role in quantum information theory.
Local operations and classical communications (LOCC) are key concepts to characterize quantum entanglement operationally.
A local operation is a measurement process performed by local observers (or a local unitary evolution), and classical communication represents the exchange of classical information, such as measurement outcomes, among observers. 
A theorem stating that LOCC cannot generate quantum entanglement \cite{Horodecki} indicates that nonlocal quantum operations are required to generate entanglement.
As discussed in the BMV proposal, the detection of gravity-induced entanglement for masses shows that gravitational interaction is described by a quantum process rather than a classical process (LOCC).

As a model to realize the macroscopic superposition of masses and to test the quantum nature of gravity, an optomechanical system has received considerable attention \cite{Marshall, Balushi,Miao,CWan, Matsumura}.
In this model, mechanical mirrors with a macroscopic mass transform to a superposition state by interacting with the cavity photons.
In Ref. \cite{Balushi, Miao}, the quantum feature appearing through gravitational interactions for optomechanical systems was analyzed through perturbative and linearized approaches, respectively. 
In Ref. \cite{CWan, Matsumura}, the gravity-induced entanglement between photons though mechanical mirrors was analyzed in a nonperturbative way.
Furthermore, the effects of quantum decoherence for photons \cite{Mancini} and mechanical mirrors \cite{Bose2} were discussed through a similar model.
However, the entanglement between mechanical mirrors coupled to cavity photons has not been estimated in the previous works.

Notably, the mechanical mirrors are in a non-Gaussian state owing to the interaction with photons. 
The entanglement of Gaussian states has been well studied \cite{Duan,Simon,Mancini2,Adesso}, and is characterized by the expectation values and covariance matrix. 
However, it is difficult to capture the entanglement properties of non-Gaussian states fully because all the higher-order statistics of non-Gaussian states are required.
Several approaches toward observing the entanglement of non-Gaussian states have been proposed using the Shannon entropy \cite{Walborn,Huang}, higher-order moments \cite{Shchukin,Serafini}, and Gaussianity \cite{Hertz}.
To test the gravity-induced entanglement, we need to prepare two massive objects in a superposition state, which generally corresponds to non-Gaussian states. 
Non-Gaussian entanglement plays an important role in the analysis of the quantum nature of gravity.

In the present study, we develop entanglement criteria even for the non-Gaussian states by using higher-order moments, particularly focusing on cumulants, mainly third- and fourth-order moments.
The cumulants are quantities that characterize non-Gaussianity \cite{Dubost,Olsen}.
To examine the non-Gaussian entanglement, we first apply the formula to a cat-like state, which is a non-Gaussian entangled state, and exemplify how the entanglement can be successfully detected.
We show that the cumulant-based criterion detects entanglement, whereas the criterion based on the covariance matrix fails.
Second, we apply the criterion to the system of the two gravitating mirrors coupled to cavity photons to demonstrate the entanglement behavior between the mirrors.
The state of the mirrors is the superposition of Gaussian states owing to the interaction with the cavity photons.
We show that the inclusion of the third- and fourth-order moments, particularly the cumulants, to characterize the non-Gaussianity is necessary to evaluate the entanglement between the mirrors properly.

The remainder of this paper is organized as follows.
In Sec. 2, we derive the entanglement criteria for the non-Gaussian states based on higher-order moments, particularly focusing on cumulants.
In Sec. 3, we evaluate the entanglement in a cat-like state as a first example of a non-Gaussian state.
We show that the entanglement can be detected by the cumulant-based criterion, whereas the criterion for 
the Gaussian states with up to second-order moments fails to detect.
In Sec. 4, we introduce the optomechanical system and examine the entanglement between the mechanical mirrors.
We also discuss the role of the cumulants in the entanglement in non-Gaussian states.
In Sec. 5, we discuss the detectability of entanglement, focusing on the Wigner function.
We also derive a lower bound of the minimum eigenvalue that characterizes the entanglement in the formula.
Sec. 6 presents the summary and conclusions.
In Appendix A, we describe the relation between the cumulants and the moments.
In Appendix B, we describe the operation of the partial transposition of the cumulants.
In Appendix C, we derive the cumulant for the optomechanical system of interest.

\section{\label{sec:level2}
Entanglement criterion based on cumulant
}
In this section, we develop a cumulant-based condition to evaluate the entanglement of non-Gaussian states.
We first introduce the positive partial transposition (PPT) criterion 
to determine whether a given quantum system is entangled \cite{Peres,Horodecki1996}.
A quantum system composed of two subsystems, A and B, is separable (not entangled) if the density matrix of the entire system is given by
\begin{align}
\label{sep}
\rho
&=\sum_{j}p_{j}\rho_{j}^\text{A}\otimes\rho_{j}^\text{B},
\end{align}
where $\rho_{j}^\text{A}$ and $\rho_{j}^\text{B}$ are the density matrices of each subsystem, and $p_{j}$ is a probability satisfying $\sum_{j}p_{j}=1$.
The partial transposition of the separable density matrix 
$\rho$ is
\begin{align}
\label{seppt}
\rho^{\text{T}_\text{B}}
&=\sum_{j}p_{j}\rho_{j}^\text{A}\otimes(\rho_{j}^\text{B})^\text{T},
\end{align}
which is non-negative because the transposed density matrix $(\rho^\text{B}_j)^\text{T}$ is non-negative.
Hence, a quantum system is entangled if the partial transposition of its density matrix has at least one negative eigenvalue.
This condition is known as the PPT criterion \cite{Peres, Horodecki1996}.

We rewrote the PPT criterion to obtain the cumulant-based condition for entanglement. 
The condition that the partial transposition $\rho^{\text{T}_\text{B}} $
of a density matrix is non-negative is equivalent to the inequality
\begin{align}
\label{pptrho}
\text{Tr}[\rho^{\text{T}_\text{B}}\hat{f}^{\dagger}\hat{f}]
&\ge0
\end{align}
for an arbitrary operator 
$\hat{f}$. 
Hence, the state is entangled if there exists an operator 
$\hat{f}$ that leads to 
$\text{Tr}[\rho^{\text{T}_\text{B}}\hat{f}^{\dagger}\hat{f}]<0$.
In Ref. \cite{Shchukin}, the entanglement criterion for a continuous variable (CV) system was derived by choosing the operator $\hat{f}$ for a polynomial of creation and annihilation operators arranged in normal order. 

In the following, to relate the non-Gaussianity and quantum entanglement for a CV system, we provide a formula for the inequality \eqref{pptrho} while clarifying the role of cumulants.
For this purpose, we choose the operator 
$\hat{f}$ as
\begin{align}
\label{f}
\hat{f}
&=z_{i}(\hat{r}_{i}-\text{Tr}[\rho^{\text{T}_\text{B}}\hat{r}_{i}])+\zeta_{ij}(\hat{r}_{i}-\text{Tr}[\rho^{\text{T}_\text{B}}\hat{r}_{i}])(\hat{r}_{j}-\text{Tr}[\rho^{\text{T}_\text{B}}\hat{r}_{j}]),\quad (i,j=1,2,3,4)
\end{align}
where 
$\rho$ is the density matrix of the CV system, $\hat{\bm{r}}=(\hat{x}_\text{A},\hat{p}_\text{A},\hat{x}_\text{B},\hat{p}_\text{B})^\text{T}$ are canonical operators, and 
$z_{i}$ and $\zeta_{ij}$ are complex numbers.  
The canonical commutation relations are expressed by $[\hat{r}_i,\hat{r}_j]=i\Omega_{ij}$, where $\Omega$ is an antisymmetric matrix given by
\begin{align}
\label{Ome}
\Omega
=
\left(
\begin{array}{cccc}
0&1&0&0\\
-1&0&0&0\\
0&0&0&1\\
0&0&-1&0
\end{array}
\right)~.
\end{align}
The left-hand side of the inequality \eqref{pptrho} is
\begin{align}
\label{ptineq}
\text{Tr}[\hat{\rho}^{\text{T}_\text{B}}\hat{f}^{\dagger}\hat{f}]
&
=
z_{i}^{*}z_{k}
\text{Tr}
[
\hat{\rho}^{\text{T}_\text{B}}
\tilde{\Delta}\hat{r}_{i}
\tilde{\Delta}\hat{r}_{k}
]
+
z_{i}^{*}\zeta_{kl}
\text{Tr}
[
\hat{\rho}^{\text{T}_\text{B}}
\tilde{\Delta}\hat{r}_{i}
\tilde{\Delta}\hat{r}_{k} \tilde{\Delta}\hat{r}_{l}
]
\notag
\\
&
\quad
+
z_{k}\zeta_{ij}^{*}
\text{Tr}
[
\hat{\rho}^{\text{T}_\text{B}}
\tilde{\Delta}\hat{r}_{j} 
\tilde{\Delta}\hat{r}_{i} \tilde{\Delta}\hat{r}_{k}
]
+
\zeta_{ij}^{*}\zeta_{kl}
\text{Tr}
[
\hat{\rho}^{\text{T}_\text{B}}
\tilde{\Delta}\hat{r}_{j} 
\tilde{\Delta}\hat{r}_{i} 
\tilde{\Delta}\hat{r}_{k}
\tilde{\Delta}\hat{r}_{l}
],
\end{align}
where we define the operator $\tilde{\Delta}\hat{r}_{i}=\hat{r}_{i}-\text{Tr}[\rho^{\text{T}_\text{B}}\hat{r}_{i}]$.
Using the vectors $\bm{z}=(z_{1},z_{2},z_{3},z_{4})^{\text{T}}$ and $\bm{\zeta}=(\zeta_{11},\zeta_{12},\cdots,\zeta_{44})^{\text{T}}$,
the right-hand side of \eqref{ptineq} is expressed in matrix form as
\begin{align}
\label{20mat}
\text{Tr}[\rho^{\text{T}_\text{B}}\hat{f}^{\dagger}\hat{f}]
&=
\left(
\begin{array}{c}
\bm{z}^{*}\\
\bm{\zeta}^{*}
\end{array}
\right)^\text{T}
\left(
\begin{array}{cc}
A&B\\
B^{\dagger}&D
\end{array}
\right)
\left(
\begin{array}{c}
\bm{z}\\
\bm{\zeta}
\end{array}
\right),
\end{align} 
where $A$, $B$, and $D$ are $4\times4$, $4\times16$, and $16\times16$ matrices, respectively.
Each matrix is
\begin{align}
\label{A}
A
&=\left(
\begin{array}{ccc}
A_{11}&\cdots&A_{14}\\
\vdots&\ddots&\vdots\\
A_{41}&\cdots&A_{44}
\end{array}
\right),\quad 
A_{ij}=
\frac{1}{2}(\sigma_{ij}^{\text{T}_\text{B}}+i\Omega_{ij}),\\
\label{B}
B
&=\left(
\begin{array}{cccc}
B_{1,11}&B_{1,12}&\cdots&B_{1,44}\\
\vdots&\vdots&\ddots&\vdots\\
B_{4,11}&B_{4,12}&\cdots&B_{4,44}
\end{array}
\right),\quad B_{i,jk}=\kappa_{3,ijk}^{\text{T}_\text{B}},\\
\label{D}
D
&=\left(
\begin{array}{cccc}
D_{11,11}&D_{11,12}&\cdots&D_{11,44}\\
D_{12,11}&D_{12,12}&\cdots&D_{12,44}\\
\vdots&\vdots&\ddots&\vdots\\
D_{44,11}&D_{44,12}&\cdots&D_{44,44}
\end{array}
\right),\quad
D_{ij,k\ell}
=\kappa_{4,ijk\ell}^{\text{T}_\text{B}}
+A_{ij}^{*}A_{k\ell}+A_{ik}A_{j\ell}+A_{i\ell}A_{jk},
\end{align}
where $\bm{\kappa}_{n}$ is the $n$th-order cumulant given by
\begin{align}
\label{ncum}
\kappa_{n,j_{1}j_{2},\cdots,j_{n}}
&=\left[\left(i\Omega_{j_{1}k_{1}}\frac{\partial}{\partial r_{k_{1}}}\right)\cdots\left(i\Omega_{j_{n}k_{n}}\frac{\partial}{\partial r_{k_{n}}}\right)\text{ln}\chi(\bm{r})\right]\biggl|_{\bm{r}=0},
\end{align}
and $\chi(\bm{r})=\text{Tr}[\rho e^{i\bm{r}^\text{T}\Omega\hat{r}}]$ is the characteristic function.
Here, $\sigma$ is the second-order central moment, called the covariance matrix.
The relations among the central moments and the cumulants are presented in Appendix A.
The covariance matrix 
$\sigma^{\text{T}_\text{B}}$ and 
the higher-order cumulants
$\kappa^{\text{T}_\text{B}}_3$ and 
$\kappa^{\text{T}_\text{B}}_4$ of the partial transposition of the density matrix are derived in Appendix B.
From the PPT criterion, we observe that a density matrix 
$\rho$ is entangled if the inequality
\begin{align}
\label{PPTineq}
M\equiv
\left(
\begin{array}{cc}
A&B\\
B^{\dagger}&D
\end{array}
\right)
\ge0
\end{align}
is violated. Here, the positivity of a matrix indicates the positivity of all the eigenvalues of the matrix.
This condition can characterize the entanglement of non-Gaussian states 
by third- and fourth-order cumulants.

Let us compare the inequality \eqref{PPTineq} with 
the PPT criterion for bipartite Gaussian states, whose entanglement is determined by the covariance matrix \cite{Simon}. 
A $1\times 1$ mode Gaussian state is entangled if and only if the inequality
\begin{align}
\label{gauss}
\sigma^{\text{T}_\text{B}}+i\Omega\ge0,
\end{align}
is violated.
This inequality \eqref{gauss} is equivalent to $A\ge0$.
The matrices $B$ and $D$, which contain the third- and fourth-order cumulants, characterize the impact of non-Gaussian features on quantum entanglement.

\section{CAT-LIKE STATE}
In this section, we present an example in which non-Gaussianity plays an important role in estimating quantum entanglement. 
We consider a two-particle state in a superposition of two coherent states 
$\ket{\alpha}_1 \ket{\alpha}_2$ and 
$\ket{-\alpha}_1 \ket{-\alpha}_2$
as
\begin{align}
\label{catlike}
\ket{\psi}
&=\frac{\ket{\alpha}_{1}\ket{\alpha}_{2}-\ket{-\alpha}_{1}\ket{-\alpha}_{2}}{\sqrt{2-2e^{-4|\alpha|^2}}},
\end{align}
where the coherent state $\ket{\alpha}$ of a single particle with a complex number 
$\alpha$ is defined as follows:
\begin{align}
\label{alpha}
\ket{\alpha}
&=e^{-|\alpha|^2/2}\sum_{n=0}^{\infty}\frac{\alpha^{n}}{\sqrt{n!}}\ket{n}
\end{align}
in the Fock basis 
$\ket{n}$.
The two-particle state is evidently entangled. 
We detect the entanglement using a criterion based on the matrix $M$ \eqref{PPTineq}.  

\begin{figure}[t]
\includegraphics[width=7.5cm]{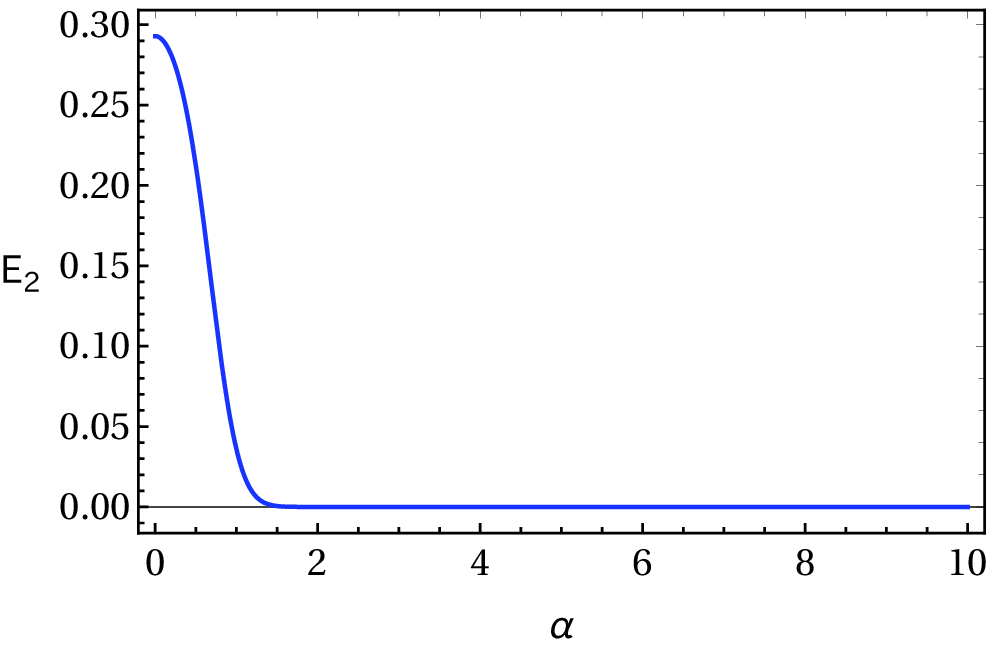}
\hspace{1cm}
\includegraphics[width=7.5cm]{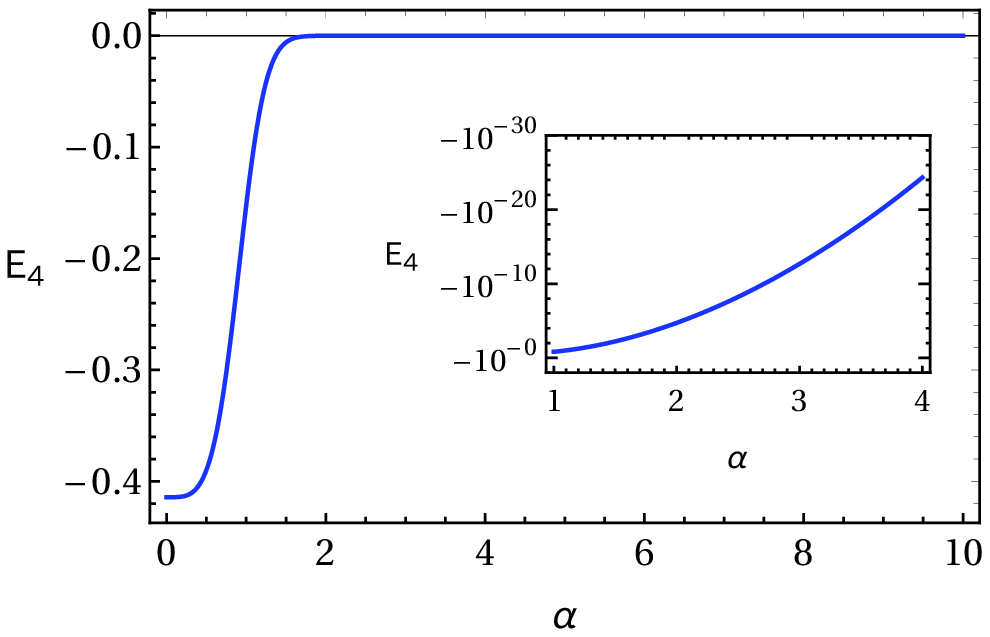}
\caption{Left: Behavior of the minimum eigenvalue $E_{2}$ of the second-order matrix $A$ as a function of $\alpha$ for the state $\ket{\psi}$ \eqref{catlike}.
$E_{2}$ is always non-negative; hence, the entanglement cannot be detected.
Right: Behavior of the minimum eigenvalue $E_{4}$ of the matrix $M$.
The entanglement of the state $\ket{\psi}$ for $\alpha\simlt2$ is detected because $E_{4}$ is negative.
The small box in the right panel shows that $E_{4}$ approaches zero from the negative region for $\alpha>1$.}
\label{fig:catlike}
\end{figure}

The characteristic function of 
$\ket{\psi}$ is obtained as
\begin{align}
\chi(\bm{r})
&=\frac{e^{-\frac{1}{4}\bm{r}^\text{T}\bm{r}}}{2-2e^{-4|\alpha|^{2}}}
\left(
e^{i\bm{\xi}^\text{T}\Omega\bm{r}}
+e^{-i\bm{\xi}^\text{T}\Omega\bm{r}}
-e^{-4|\alpha|^{2}}(e^{\bm{\xi}^\text{T}\bm{r}}
+e^{-\bm{\xi}^\text{T}\bm{r}})
\right),
\end{align}
where 
$\bm{\xi}=\sqrt{2}(\text{Re}[\alpha],\text{Im}[\alpha],\text{Re}[\alpha],\text{Im}[\alpha])^\text{T}$.
The Wigner function is defined as
\begin{align}
W(\bm{X})
&=\frac{1}{2\pi}\int d^{2}\bm{r}\chi{\bm{r}}e^{i\bm{X}^{T}\Omega\bm{r}}\notag\\
&=\frac{1}{1-e^{-4|\alpha|^{2}}}\left(
e^{-(\bm{X}+\bm{\xi})^{T}(\bm{X}+\bm{\xi})}+e^{-(\bm{X}-\bm{\xi})^{T}(\bm{X}-\bm{\xi})}
-2e^{-4|\alpha|^{2}}e^{-\bm{X}^{T}\bm{X}+\bm{\xi}^{T}\bm{\xi}}\cos(2\bm{X}^{T}\Omega\bm{\xi})
\right).
\end{align}
In general, the Wigner function $W(\bm{X})$ is not necessarily positive:
the negative values characterize the nonclassicality.
Here, we focus on the quantum entanglement by the cumulants to analyze quantumness.
Using Eq.~\eqref{ncum}, we derive the covariance matrix and the cumulants as
\begin{align}
\label{catA}
\sigma_{ij}
&
=\bm{1}_{ij}
-\frac{2\Omega_{ia}\Omega_{jb}}{1-e^{-4|\alpha|^{2}}}\Xi_{ab},\\
\label{catB}
\kappa_{3,ijk}
&=0,\\
\label{catD}
\kappa_{4,ijk\ell}
&=\frac{1}{1-e^{-4|\alpha|^{2}}}
\left\{
\xi_{i}\xi_{j}\xi_{k}\xi_{\ell}
-e^{-4|\alpha|^{2}}\Omega_{ia}\Omega_{jb}\Omega_{kc}\Omega_{\ell d}\xi_{a}\xi_{b}\xi_{c}\xi_{d}\right.\notag\\
&\left.
-\frac{\Omega_{ia}\Omega_{jb}\Omega_{kc}\Omega_{\ell d}}{1-e^{-4|\alpha|^{2}}}
\Big(
\Xi_{ab}\Xi_{cd}+\Xi_{ac}\Xi_{bd}+\Xi_{ad}\Xi_{bc}
\Big)
\right\},
\end{align}
where $\Xi_{ab}$ is given by
\begin{align}
\Xi_{ij}
&=-\Omega_{ia}\Omega_{jb}\xi_{a}\xi_{b}-e^{-4|\alpha|^{2}}\xi_{i}\xi_{j}.
\end{align}
Fig.~\ref{fig:catlike} presents the behaviors of the minimum eigenvalues 
$E_2$ and $E_4$ of the matrices $A$
and $M$, respectively.
As shown in the left panel of Fig.~\ref{fig:catlike}, the minimum eigenvalue 
$E_2$ of 
$A$ is non-negative for 
$\alpha \geq 0$; hence, the entanglement between the two particles is not detected by only the covariance matrix. 
Note that, if the matrix $A$ has a negative eigenvalue, then the inequality \eqref{PPTineq} is violated. 
In contrast, in the right panel of Fig.~\ref{fig:catlike}, we observe that the matrix $M$ has a negative eigenvalue and violates the inequality \eqref{PPTineq} for
$\alpha \simlt 2$. 
Therefore, the non-Gaussian feature characterized by the third- and 
fourth-order cumulants is important for estimating entanglement. 
The entanglement behavior in Fig. \ref{fig:catlike} for a small
$\alpha$ can be explained as follows:
For $\alpha\ll1$, the cat-like state is written as
\begin{align}
\label{catmax}
\ket{\Psi_{cl}}
&\sim \frac{2\alpha e^{|\alpha|^{2}}}{\sqrt{2-2e^{-4|\alpha|^{2}}}}(\ket{0}_{1}\ket{1}_{2}+\ket{1}_{1}\ket{0}_{2}),\notag\\
&\sim \frac{1}{\sqrt{2}}(\ket{0}_{1}\ket{1}_{2}+\ket{1}_{1}\ket{0}_{2}),\quad \alpha\in\mathbb{R}.
\end{align}
Then, the state approaches the entangled state for $\alpha\ll1$.
The criterion with up to the fourth-order cumulant detects this entanglement around $\alpha \simlt 2$, although the criterion based on the second-order moment fails to detect the entanglement.
We note that the minimum eigenvalue $E_{4}$ is a small negative value even when $\alpha\simgt2$, which is consistent with Ref. \cite{Shchukin}.

\section{Gravitating mirrors in optomechanical systems}

\begin{figure}[b]
\centering
\includegraphics[width=10.0cm]{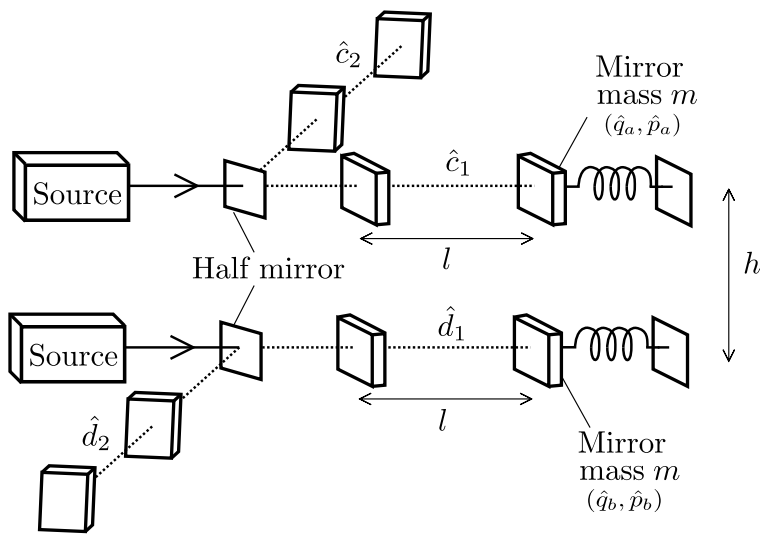}
\caption{Set up of the optomechanical system similar to the model proposed in Ref. \cite{Balushi,Miao,Matsumura}
}
\label{fig:setup}
\end{figure}

In this section, we consider an optomechanical system with two mechanical mirrors and four cavities, which is similar to the model proposed in Ref. \cite{Balushi,Miao,Matsumura}.
Each mirror with mass $m$ is placed at the end of one cavity, and the two mirrors are vertically separated by a distance $h$, as illustrated in Fig. \ref{fig:setup}.
The mirrors are coupled to photons in the superposition states and oscillate owing to the pressure of the photons.

The Hamiltonian describing the optomechanical system is given by
\begin{align}
\label{hamiltonian}
\hat{H}
&=\frac{\hbar\Omega_{m}\sqrt{1+g}}{2}(\hat{p}_{a}^{2}+\hat{q}_{a}^{2})+\frac{\hbar\Omega_{m}\sqrt{1+g}}{2}(\hat{p}_{b}^{2}+\hat{q}_{b}^{2})+\hbar\omega_{p}(\hat{c}_{1}^{\dagger}\hat{c}_{1}+\hat{c}_{2}^{\dagger}\hat{c}_{2})+\hbar\omega_{p}(\hat{d}_{1}^{\dagger}\hat{d}_{1}+\hat{d}_{2}^{\dagger}\hat{d}_{2})\notag\\
&\quad-\hbar\frac{\Lambda\Omega_{m}}{(1+g)^{1/4}}\hat{c}_{1}^{\dagger}\hat{c}_{1}\hat{q}_{a}-\hbar\frac{\Lambda\Omega_{m}}{(1+g)^{1/4}}\hat{d}_{1}^{\dagger}\hat{d}_{1}\hat{q}_{b}-\hbar\frac{g\Omega_{m}}{\sqrt{1+g}}\hat{q}_{a}\hat{q}_{b},
\end{align}
where $\hat{q}_{a}$ and $\hat{p}_{a}$ ($\hat{q}_{b}$ and $\hat{p}_{b}$) are the dimensionless canonical operators of the mechanical mirrors, and $\hat{c}_{1}$ and $\hat{c}_{1}^{\dagger}$ ($\hat{d}_{1}$ and $\hat{d}_{1}^{\dagger}$) are the creation and annihilation operators of the photons, respectively (see Refs. \cite{Balushi,Matsumura}).
The gravitational coupling $g$ and the mirror--photon coupling $\Lambda$ are given by
\begin{align}
\label{glam}
g
&=\frac{Gm}{\Omega_{m}^{2}h^{3}},\quad \Lambda=\frac{\omega_{p}}{\Omega_{m}l}\sqrt{\frac{\hbar}{2m\Omega_{m}}},
\end{align}
where $G$ is the gravitational constant, $\Omega_{m}$ and $\omega_{p}$ are the frequencies of the mirrors and photons, respectively, and $l$ is the cavity length.

We assume that the initial state of the mirrors is a coherent state given by 
\begin{align}
\label{oscini}
\ket{\phi_{m}}
&=e^{i\bm{r}'{}^{\text{T}}\Omega\hat{\bm{r}}}\ket{0}_{a}\ket{0}_{b},
\end{align}
where $\bm{r}'=(q'_{a},p'_{b},q'_{b},p'_{b})^{\text{T}}$ and $\hat{\bm{r}}=(\hat{q}_{a},\hat{p}_{b},\hat{q}_{b},\hat{p}_{b})^{\text{T}}$.
In a realistic case, the initial state of the mirrors may be a mixed state, such as the thermal state, because of the interaction with the environment.
However, we simply assume that the initial state of the mirrors is a coherent state, that is,
the same assumption as that in Ref. \cite{Balushi,Matsumura}, as we focus on the entanglement due to the non-Gaussianity.
Furthermore, the following discussion holds for the initial vacuum state, $\bm{r}'=0$.
We also assume that the initial state for the photons is
\begin{align}
\label{phoini}
\ket{\phi_{p}}
&=\frac{1}{\sqrt{2}}(\ket{0}_{c_{1}}\ket{1}_{c_2}+\ket{1}_{c_{1}}\ket{0}_{c_2})\otimes\frac{1}{\sqrt{2}}(\ket{0}_{d_{1}}\ket{1}_{d_2}+\ket{1}_{d_{1}}\ket{0}_{d_2}).
\end{align}
Then, the initial state of the entire system is $\ket{\Psi(0)}=\ket{\phi_{p}}\ket{\phi_{m}}$. 
The evolved state under the Hamiltonian in Eq. \eqref{hamiltonian} is
\begin{align}
\label{totsta}
\ket{\Psi(t)}
&=\frac{1}{2}e^{-i(\omega_{c}+\omega_{d})t}\sum_{k,\ell=0}^{1}\ket{k}_{c_{1}}\ket{1-k}_{c_{2}}\ket{\ell}_{d_{1}}\ket{1-\ell}_{d_{2}}\otimes e^{-\frac{i\hbar}{2}\hat{\bm{r}}^{\text{T}}H\hat{\bm{r}}t}\ket{\phi_{k\ell}},
\end{align}
where the Hamiltonian matrix $H$ is given by
\begin{align}
\label{H}
H
&=\Omega_{m}\left(
\begin{array}{cccc}
\sqrt{1+g}&0&-\frac{g}{\sqrt{1+g}}&0\\
0&\sqrt{1+g}&0&0\\
-\frac{g}{\sqrt{1+g}}&0&\sqrt{1+g}&0\\
0&0&0&\sqrt{1+g}
\end{array}
\right),
\end{align}
and the state of the mirrors is
\begin{align}
\label{phikl}
\ket{\phi_{k\ell}}
&=e^{
i\bm{j}_{k\ell}^{\text{T}} H^{-1}\Omega[\bm{1}-e^{t\Omega H}]
\hat{\bm{r}}
+
\frac{i}{2}
\bm{j}_{k\ell}^{\text{T}}
\left(tH^{-1}-[\bm{1}-e^{t\Omega H}] H^{-1}\Omega H^{-1}\right)
\bm{j}_{k\ell}
}
\ket{\phi_{m}},
\end{align}
where 
$\bm{j}_{k\ell}=(k\Lambda\Omega_{m}(1+g)^{-\frac{1}{4}},0,\ell\Lambda\Omega_{m}(1+g)^{-\frac{1}{4}},0)^{\text{T}}$.

We focus on the gravity-induced entanglement between the mechanical mirrors. 
By tracing the photons in the total density matrix $\rho(t)=\ket{\Psi(t)}\bra{\Psi(t)}$, we obtain the reduced density matrix of the mirrors,
\begin{align}
\label{rhoosc}
\rho_{m}
&
=\frac{1}{4}
\sum_{k,\ell=0}^{1}
\rho_{m,kl},
\quad
\rho_{m,kl}=
e^{-\frac{i}{2}\hat{\bm{r}}^{\text{T}}H\hat{\bm{r}}t}
\ket{\phi_{k\ell}}\bra{\phi_{k\ell}}
e^{\frac{i}{2}\hat{\bm{r}}^{\text{T}}H\hat{\bm{r}}t},
\end{align}
and the characteristic function $\chi(\bm{r})=\text{Tr}[\rho_m e^{i\bm{r}^{\text{T}}\Omega\hat{\bm{r}}}]$ is given as
\begin{equation}
\label{cha}
\chi(\bm{r})
=
\frac{1}{4}
\sum_{k,\ell=0}^{1}
\chi_{k\ell}(\bm{r}),
\quad 
\chi_{k\ell}(\bm{r})
=
e^{-\frac{1}{4}\bm{r}^{\text{T}}S^{-1\text{T}}S^{-1}\bm{r}
-i\bm{r}^{\text{T}}\Omega S(\bm{j}'_{k\ell}+\bm{r}')},
\end{equation}
where 
$\chi_{k\ell}(\bm{r})$ is the characteristic function of the Gaussian state $e^{-\frac{i}{2}\hat{\bm{r}}^{\text{T}}H\hat{\bm{r}}t}
\ket{\phi_{k\ell}}$,
$\bm{j}'_{k\ell}=H^{-1}(1-e^{-tH\Omega})\bm{j}_{k \ell}$ and
$S=e^{\Omega Ht}$ is the symplectic matrix that satisfies 
$S\Omega S^{\text{T}}=\Omega$.
The characteristic function 
$\chi(\bm{r})$
is the sum of four Gaussian states, which leads to a non-Gaussian state, except for the case where $\Lambda=0$.
This non-Gaussian state originates from the spatial superposition of mirrors, which is induced by optomechanical interactions
proportional to the cubic terms $\hat c_1^\dagger \hat c_1 \hat q_a$
and $\hat d_1^\dagger \hat d_1 \hat q_b$ in the Hamiltonian, Eqs.~\eqref{hamiltonian}: 
We calculate the matrices $A$, $B$, and $D$ (see Appendix C for details) and demonstrate the entanglement between the mirrors owing to gravity.

\begin{figure}[H]
\begin{minipage}[t]{1\linewidth}
~
\includegraphics[width=7.5cm]{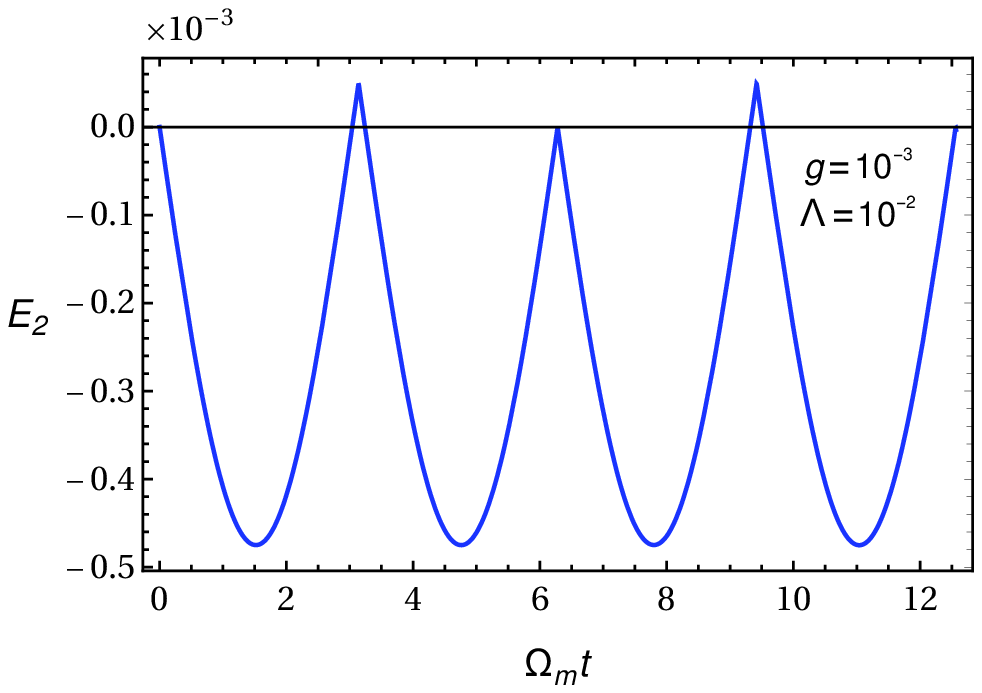}
\hfill
\includegraphics[width=7.5cm]{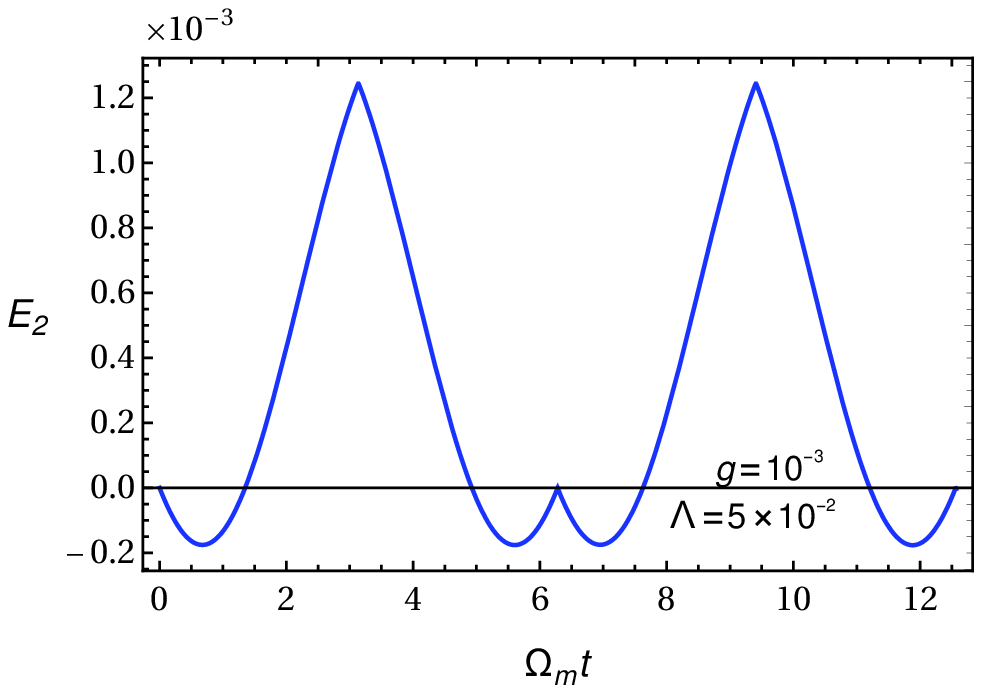}
~
\end{minipage}
\begin{minipage}[t]{1\linewidth}
~
\includegraphics[width=7.5cm]{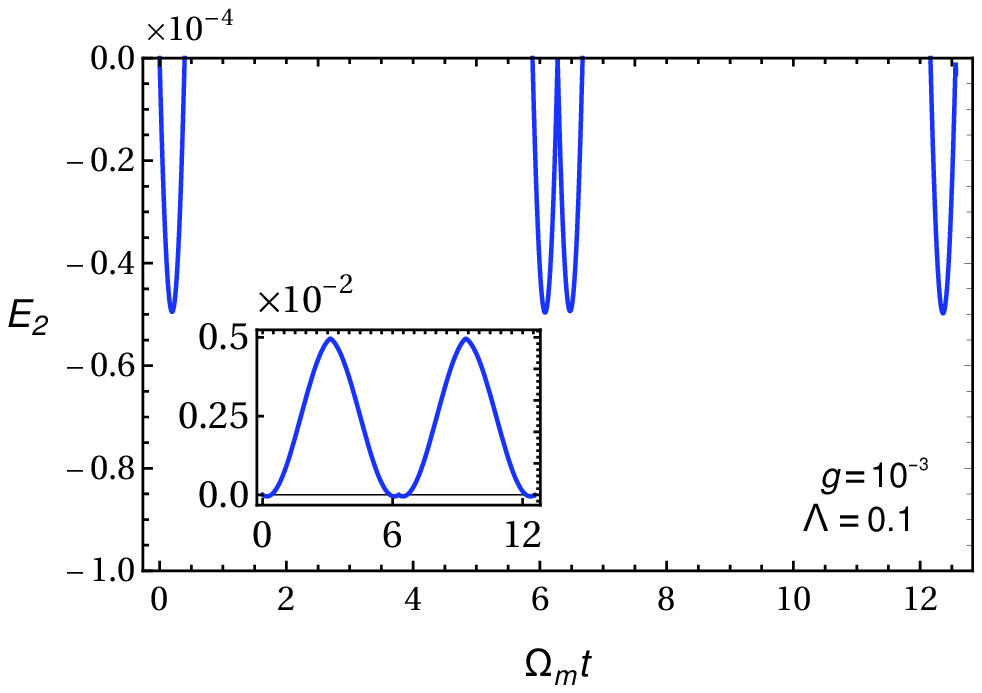}
\hfill
\includegraphics[width=7.5cm]{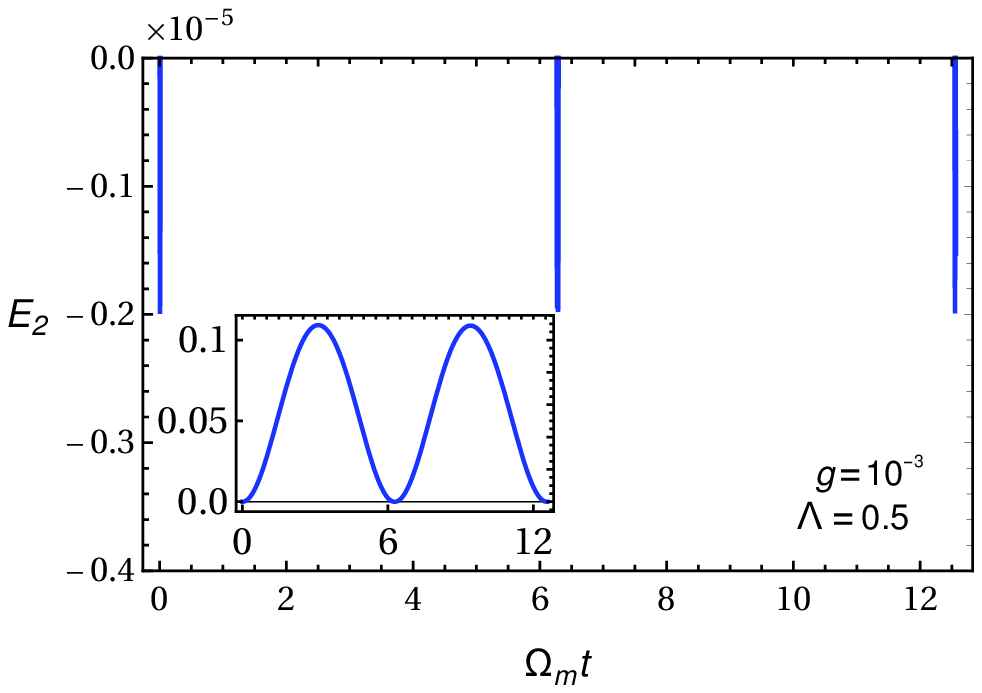}
~
\end{minipage}
\begin{minipage}[t]{1\linewidth}
~
\includegraphics[width=7.5cm]{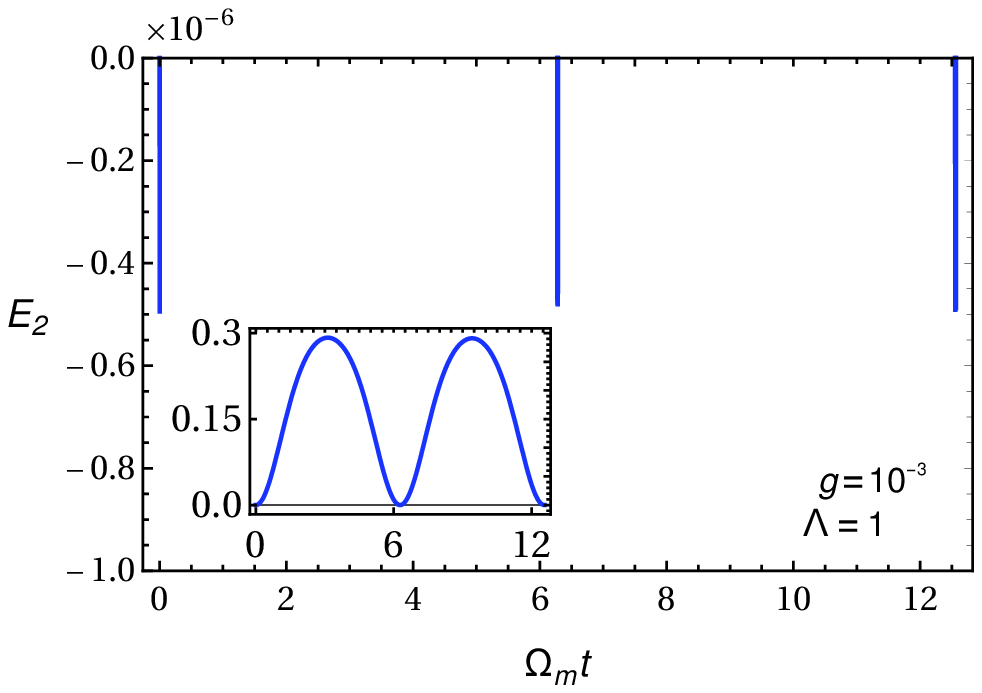}
\hfill
\includegraphics[width=7.5cm]{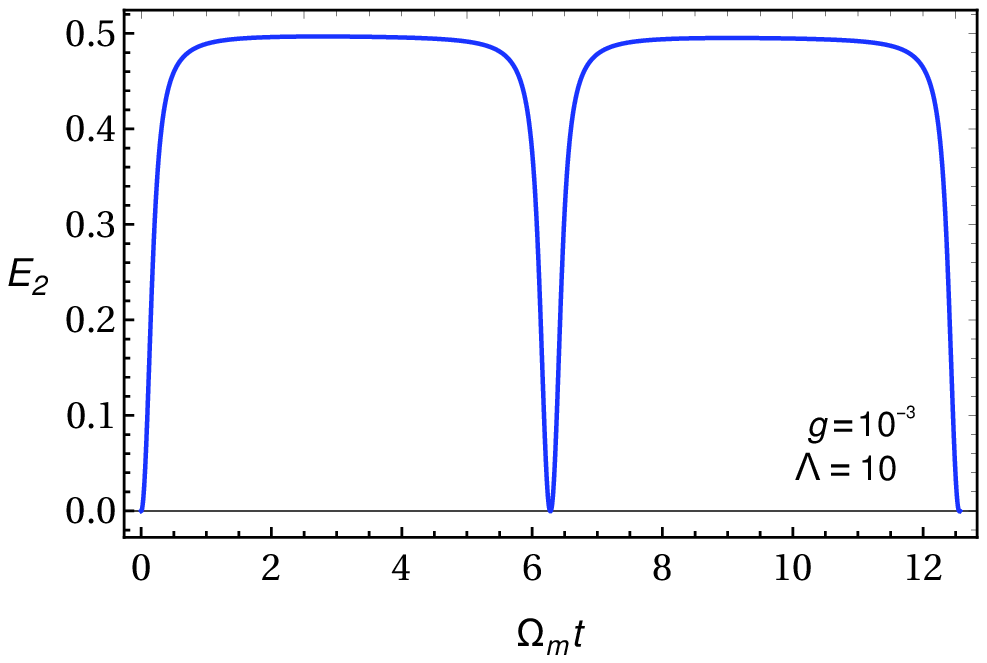}
~
\end{minipage}
\caption{ 
Behavior of the minimum eigenvalue
$E_2$ of the matrix $A$ as a function of the dimensionless time $\Omega_{m}t$.
The gravitational coupling constant is fixed as $g=10^{-3}$, and the mirror--photon coupling $\Lambda$ 
varies, whose value is noted on each panel.
The panels for $\Lambda=0.1$, $0.5$, and $1$ show only the regions with negative eigenvalues, and the overall behavior is shown in the smaller box in each panel.
When the coupling $\Lambda$ is small, the entanglement is detected because the state is almost Gaussian.
However, for the coupling $\Lambda\simgt0.1$, 
the entanglement is hardly detected by the criterion with the covariance matrix.}
\label{fig:coveig}
\end{figure}

\begin{figure}[H]
\begin{minipage}[t]{1\linewidth}
~
\includegraphics[width=7.5cm]{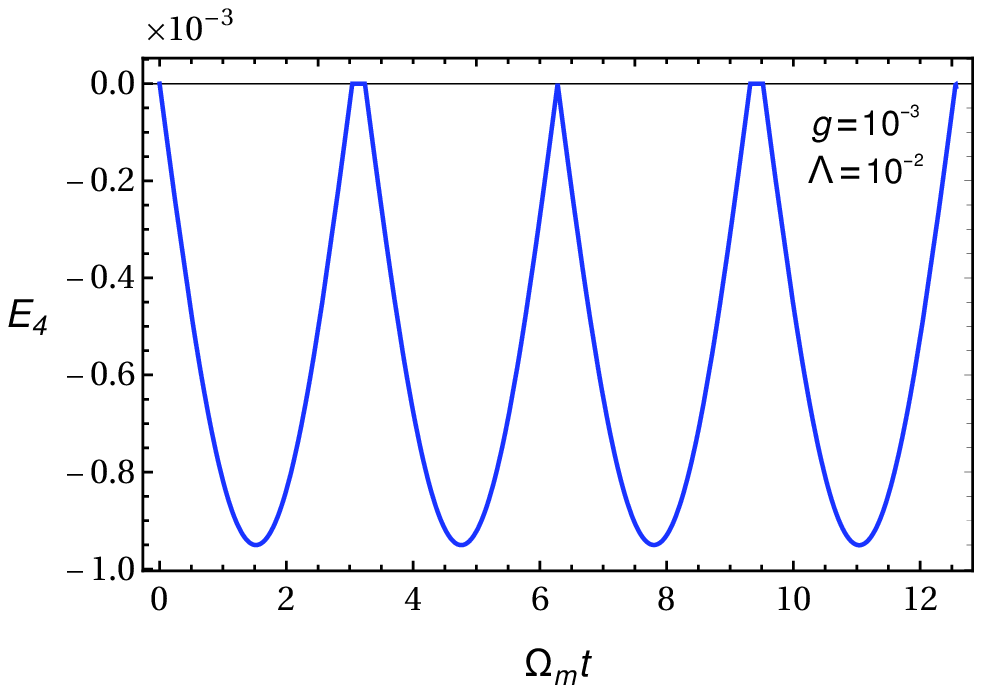}
\hfill
\includegraphics[width=7.5cm]{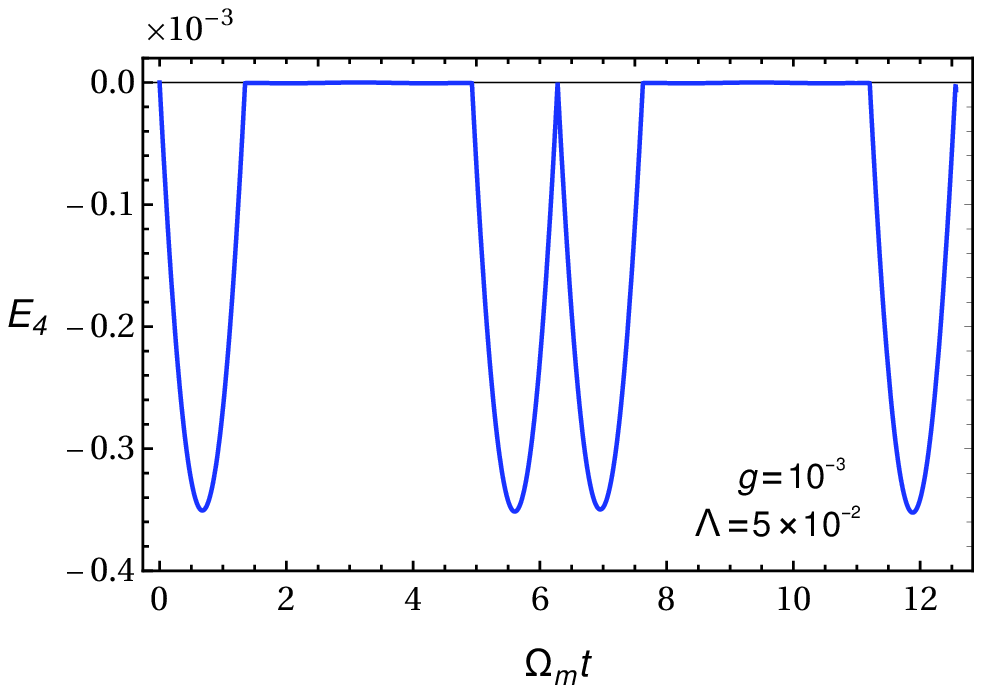}
~
\end{minipage}
\begin{minipage}[t]{1\linewidth}
~
\includegraphics[width=7.5cm]{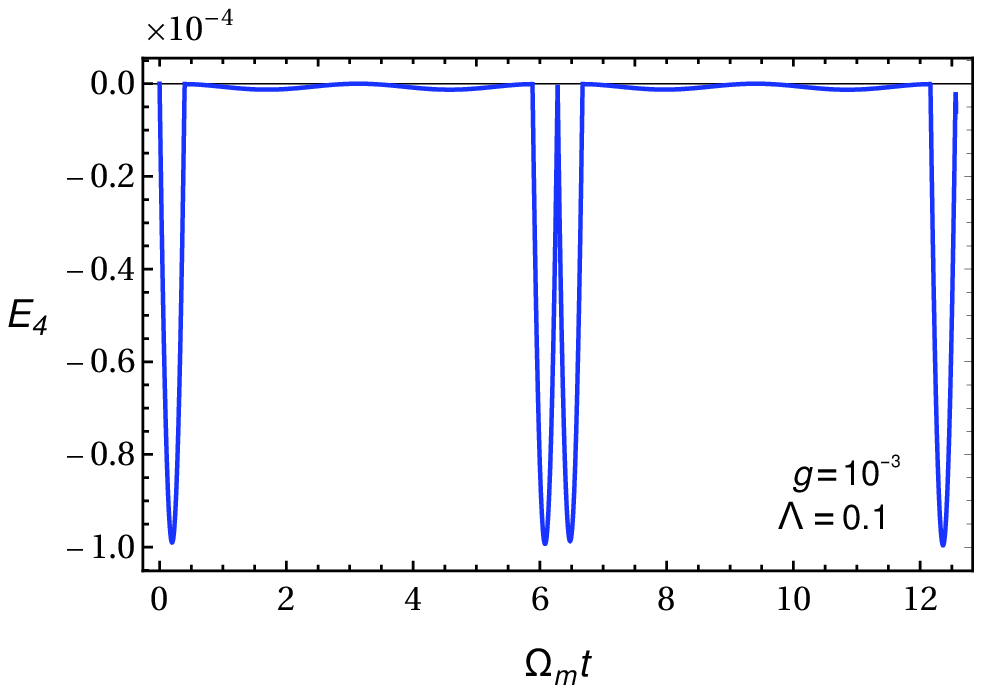}
\hfill
\includegraphics[width=7.5cm]{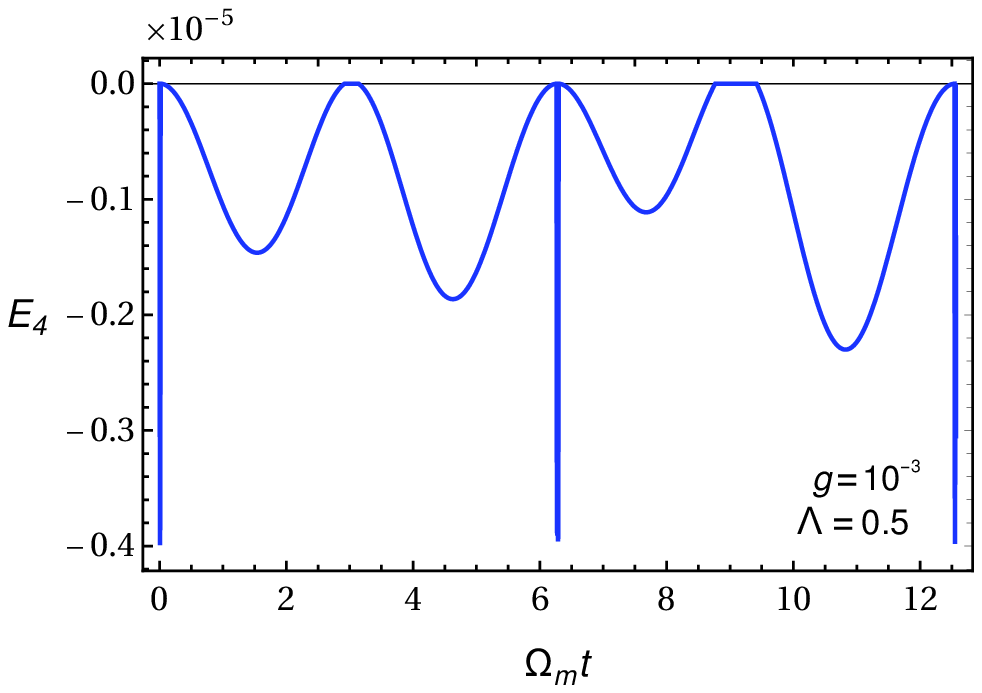}
~
\end{minipage}
\begin{minipage}[t]{1\linewidth}
~
\includegraphics[width=7.5cm]{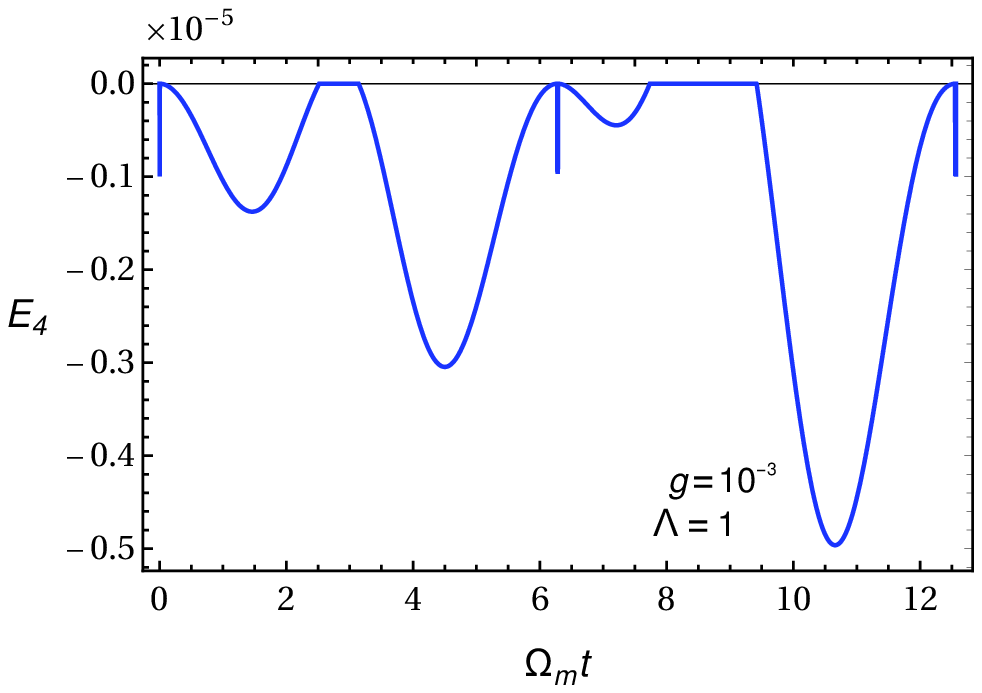}
\hfill
\includegraphics[width=7.5cm]{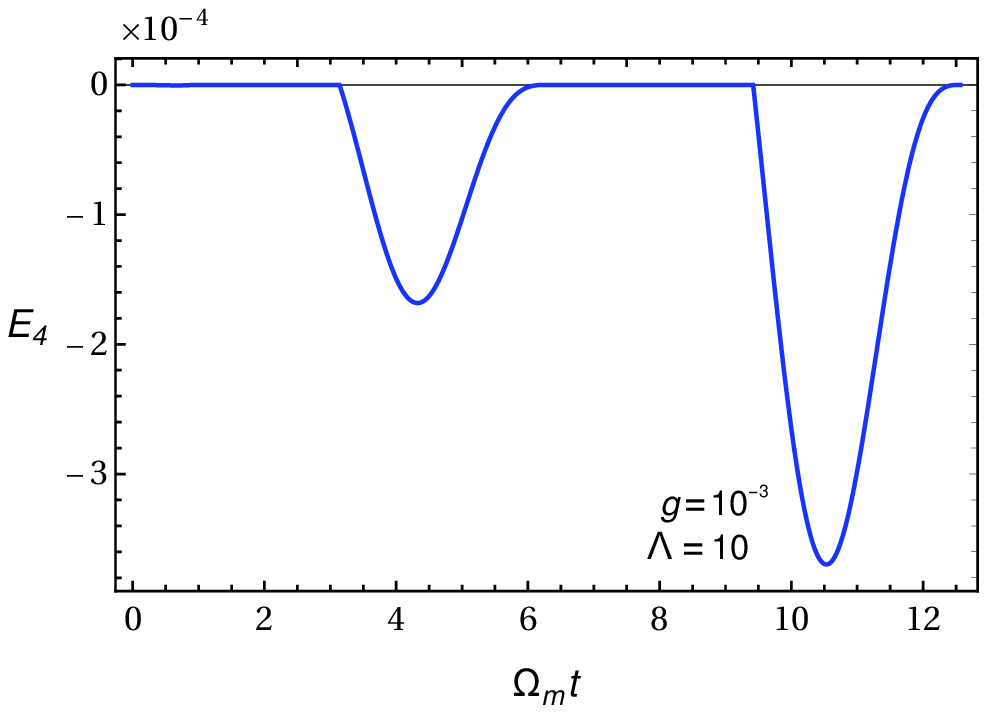}
~
\end{minipage}
\caption{
Behavior of the minimum eigenvalue
$E_4$ of the matrix $M$ as a function of the dimensionless time $\Omega_{m}t$ for the same parameters, as shown in Fig.~\ref{fig:coveig}.
This criterion using up to the fourth-order cumulant significantly improves the detection of the entanglement even when $\Lambda$ is large, i.e., $\Lambda\simgt0.5$, where the non-Gaussian feature is substantial.}
\label{fig:cumeig}
\end{figure}

Figs.~\ref{fig:coveig} and \ref{fig:cumeig} demonstrate the behaviors of the minimum eigenvalues 
$E_2$ and $E_4$ of the matrices $A$ and $M$, respectively.
Here, the gravitational coupling is fixed as
$g=10^{-3}$ and the mirror--photon coupling 
$\Lambda$ varies.
When the coupling 
$\Lambda$ is small, 
the entanglement is observed in either way using the matrix 
$A$ or 
$M$, which are composed of the covariance matrix
and up to the fourth-order cumulant, respectively.
This is because the state is almost identical to the Gaussian states when $\Lambda$ is sufficiently small.
On the other hand, as the coupling $\Lambda$ increases, the region of entanglement detected by using the matrix $A$ becomes small (see Fig. \ref{fig:coveig}). 
However, the negativity of the matrix $M$ 
can properly capture the entanglement between mirrors. 
Hence, 
the third- and fourth-order cumulants included in the matrix $M$ characterize the entanglement originating from non-Gaussianity.

\begin{figure}[t]
\begin{minipage}[t]{1\linewidth}
\quad
\includegraphics[width=7.5cm]{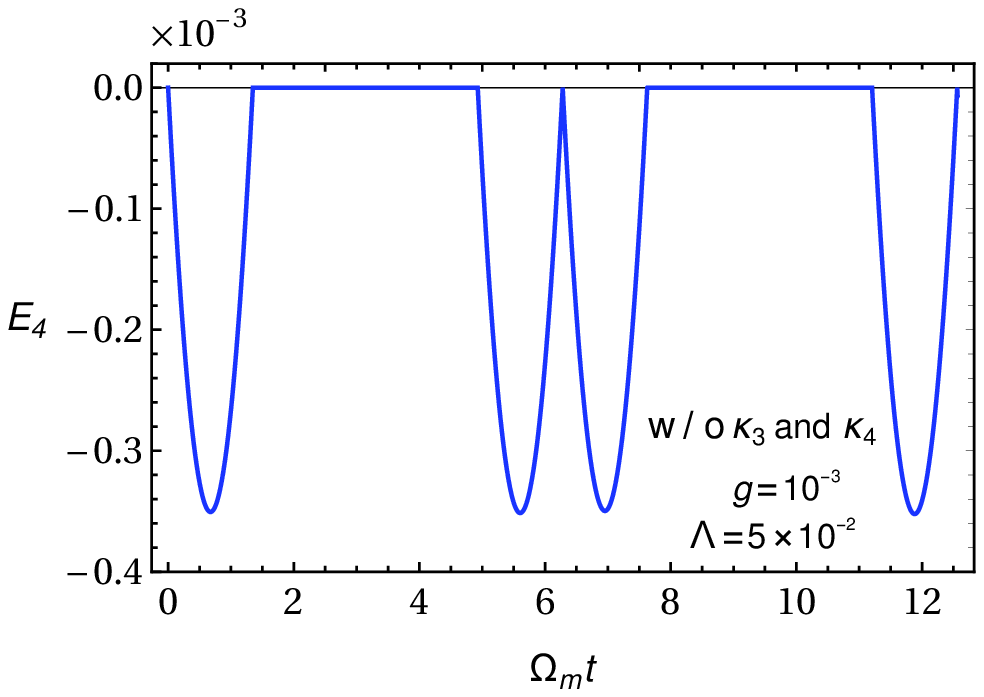}
\hfill
\includegraphics[width=7.5cm]{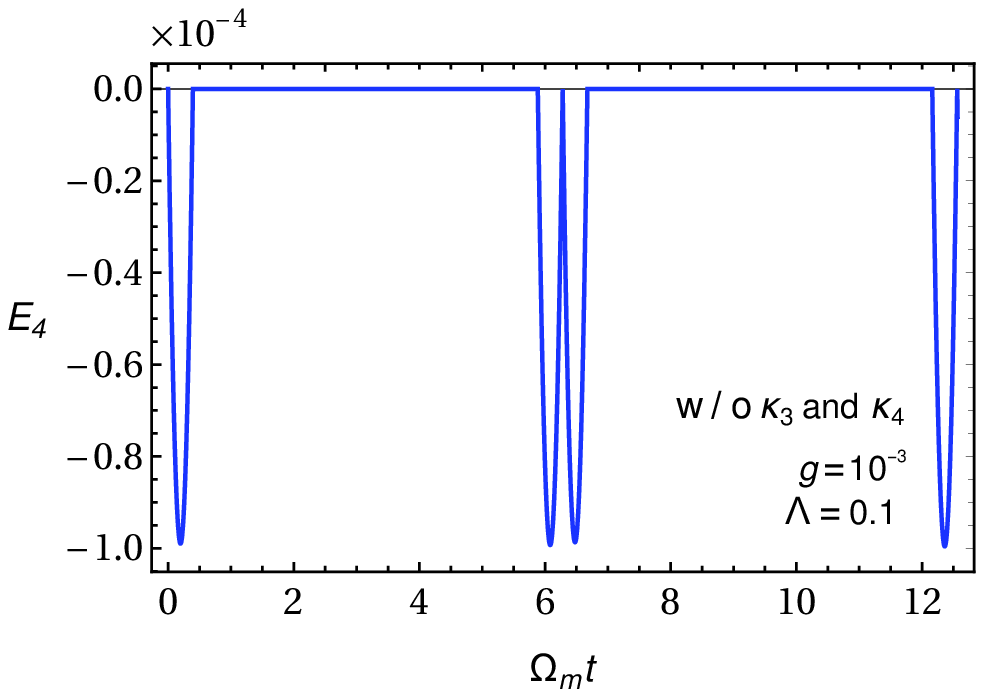}
\quad
\end{minipage}
\begin{minipage}[t]{1\linewidth}
\quad
\includegraphics[width=7.5cm]{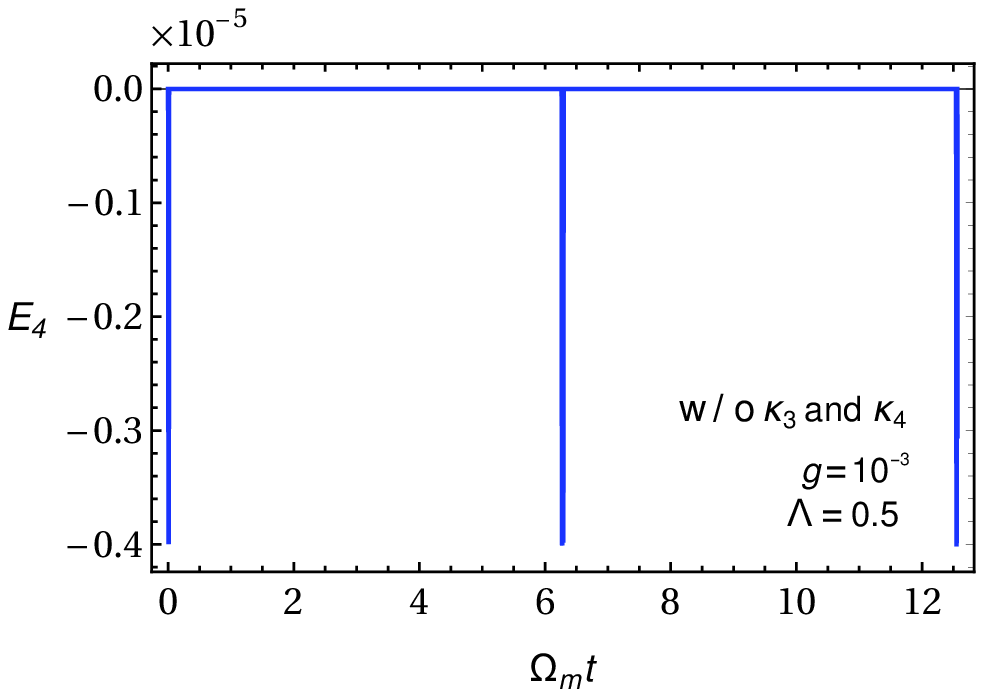}
\hfill
\includegraphics[width=7.5cm]{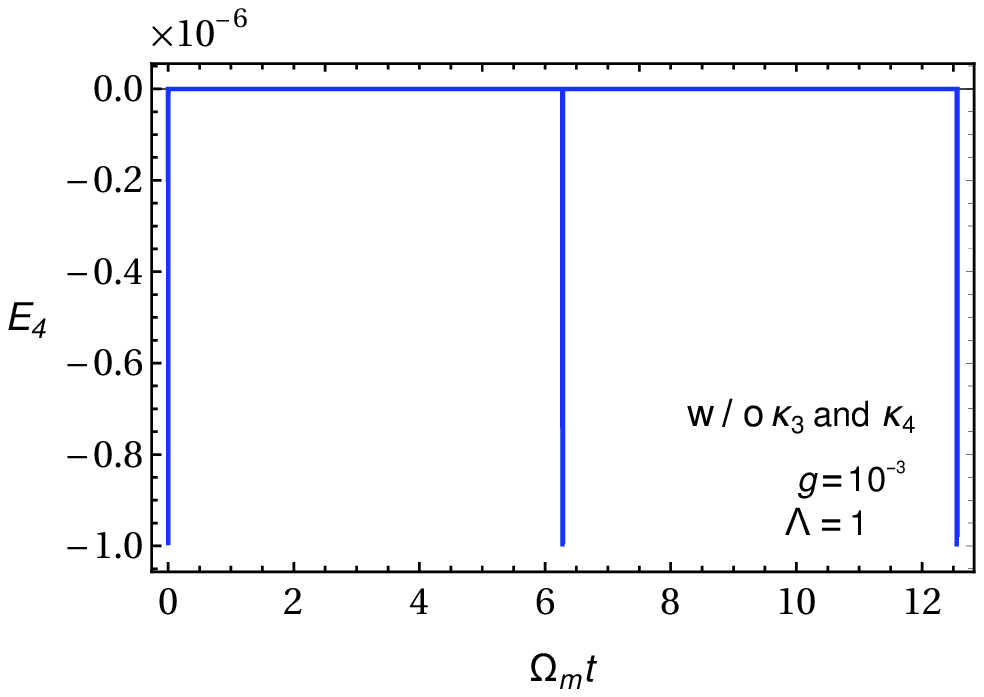}
\quad
\end{minipage}
\caption{ 
Behavior of the minimum eigenvalue
$E_4$ of the matrix $M$ without the third- and fourth-order cumulants.
The behavior is similar to that shown in Fig.~\ref{fig:cumeig}
}
\label{fig:onlycov}
\end{figure}

To clarify the role of the cumulant in detecting the quantum entanglement of the non-Gaussian state, 
we evaluate the negativity with the matrix M by assigning   
\begin{eqnarray*}
&&A_{ij}
=\frac{1}{2}(\sigma_{ij}^{\text{T}_{\text{B}}}+i\Omega_{ij}),\notag\\
&&B_{i,jk}
=0,\notag\\
&&D_{ij,kl}
=A_{ij}^{*}A_{kl}+A_{ik}A_{jl}+A_{il}A_{jk}, \notag
\end{eqnarray*}
where the third- and fourth-order cumulants were omitted. 
In this case, the matrix $M$ is determined only by the covariance matrix.
Fig.~\ref{fig:onlycov} shows the minimum eigenvalue 
$E_4$ of the matrix $M$ 
without the third- and fourth-order cumulants for the case $g=10^{-3}$.
A comparison of the panels in Fig.~\ref{fig:onlycov} with those in Fig.~\ref{fig:coveig} shows similar features. However, the behaviors are different from those shown in Fig.~\ref{fig:cumeig}.
This demonstrates that the third- and fourth-order cumulants are essential for detecting the entanglement of the gravitating mirrors.

\section{Discussions}
\begin{figure}[t]
\begin{minipage}[t]{0.48\linewidth}
\centering
\includegraphics[width=7.5cm]{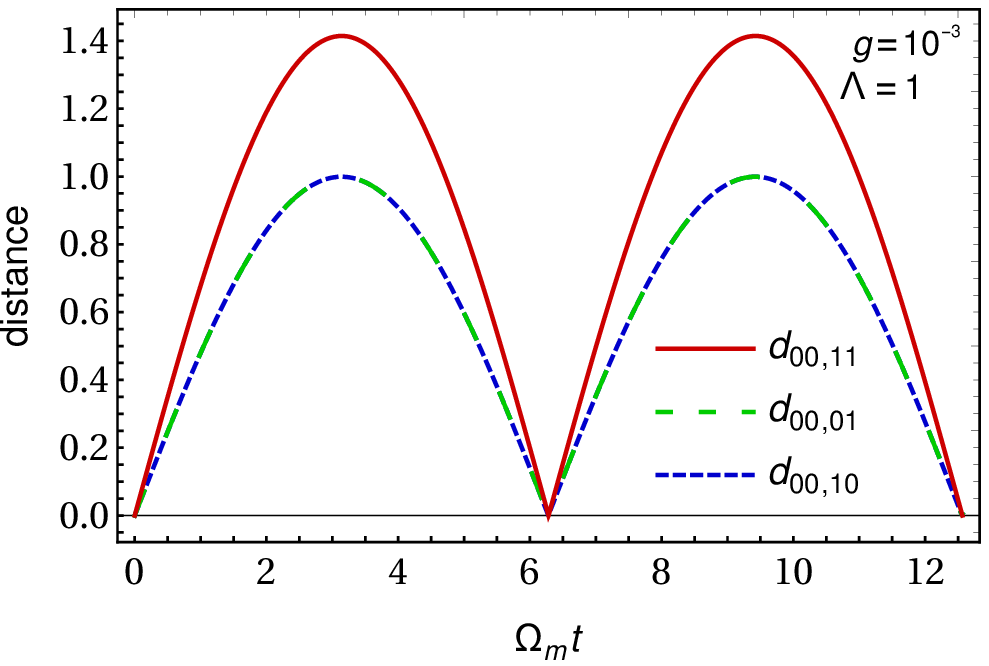}
\end{minipage}
\begin{minipage}[t]{0.48\linewidth}
\centering
\includegraphics[width=7.5cm]{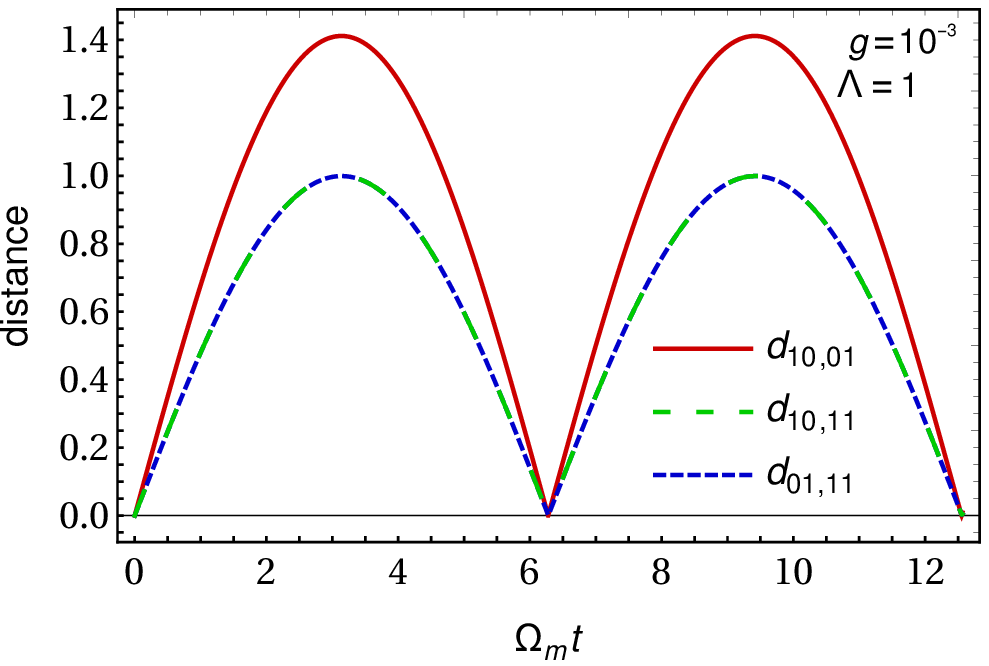}
\end{minipage}
\caption{
Behavior of the distance between the two centers of the Gaussian distribution 
defined by Eq.~(\ref{defdistance}) for $g=10^{-3}$ and $\Lambda=1$.
The state is almost Gaussian when the centers of each distribution are close.
In contrast, the state is different from the Gaussian when the 
centers of each distribution are farthest apart at $\Omega_m t=\pi, 3\pi$.}
\label{fig:wigner}
\end{figure}

In the previous section, we demonstrated the usefulness of the third- and fourth-order 
moments to detect the entanglement of non-Gaussian states. 
However, the entanglement criterion is not always appropriate. 
As shown in Fig.~\ref{fig:cumeig}, the insensitive region for the criterion is large for a large optomechanical coupling $\Lambda$. 
This occurs when 
the state differs from the Gaussian states significantly.
To demonstrate this, we use the Wigner function defined by
\begin{equation}
W(\bm{X})
=\frac{1}{4} \sum_{k,\ell=0}^{1}W_{k\ell}(\bm{X}), 
\end{equation}
with 
\begin{eqnarray}
W_{k\ell} (\bm{X})
=\frac{1}{2\pi}\int d^{2}\bm{r}\chi_{k\ell} (\bm{r})e^{i\bm{X}^{\text{T}}\Omega\bm{r}}
=\frac{\pi}{2}e^{-(\bm{X}+S(\bm{j}_{kl}'+\bm{r}'))^{\text{T}}S^{-1\text{T}}S^{-1}(\bm{X}+S(\bm{j}_{kl}'+\bm{r}'))},
\end{eqnarray}
where $\bm{X}=(Q_{a},P_{a},Q_{b},P_{b})^{\text{T}}$.
We note that the entanglement between the gravitating mirrors appears even though the Wigner function is always non-negative, which is in contrast to the cat-like state.
The center of each Gaussian distribution $W_{k\ell} (\bm{X})$
in the phase space is $\bm{X}_{kl}=-S(\bm{j}'_{kl}+\bm{r}')$, and its variance is $SS^{\text{T}}$, whose determinant equals 1.
Here, we define the distance between the two centers of the Gaussian distribution as
\begin{align}
d_{ij,kl}
&=|\bm{X}_{ij}-\bm{X}_{kl}|/\sqrt{\Sigma},
\label{defdistance}
\end{align}
where $\Sigma$ is the sum of the eigenvalues of the variance $S^{-1\text{T}}S^{-1}$.
Fig.~\ref{fig:wigner} shows the behavior of the distance for the model with $g=10^{-3}$ and $\Lambda=1$.

Compared with the left bottom panel in Fig.~\ref{fig:cumeig} with
the same parameters $g=10^{-3}$ and $\Lambda=1$, the 
entanglement is difficult to detect when the centers of each distribution are far from each other at $\Omega_mt=\pi, 3\pi$.
Hence, we infer that the entanglement criterion based on up to the fourth-order cumulant is effective for detecting the entanglement when the deviation from the Gaussian state is not large.
The validity of this conclusion is apparent for the cat-like state described in Sec. 3. 
In the right panel of Fig.~\ref{fig:catlike}, the value of the negativity 
is small for the large values of $\alpha\simgt 2$, 
where the non-Gaussian feature is significant. Here, the
entanglement of the non-Gaussian cat-like state is not well detected.

\begin{figure}[t]
\centering
\includegraphics[width=7.5cm]{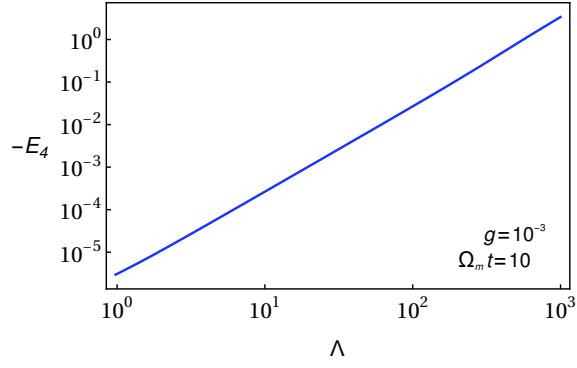}
\caption{
Behavior of the negative value of the minimum eigenvalue as a function of the coupling $\Lambda$ for $g=10^{-3}$ and $\Omega_{m}t=10$.
The minimum eigenvalue decreases in proportion to $\Lambda^{2}$.
}
\label{fig:LL}
\end{figure}

We now analyze the behavior of the minimum eigenvalue for a large $\Lambda$.
As shown in Fig.~\ref{fig:cumeig}, the cumulant-based condition enables the detection of entanglement even when the interaction between the mirrors and the photons
is sufficiently strong, $\Lambda\simgt1$, and the non-Gaussianity is significant.
Here, we show that there is a lower bound of the minimum eigenvalue of the matrix $M$.
From Eq.~\eqref{f} and Eq.~\eqref{rhoosc}, we have
\begin{align}
\label{bound}
\text{Tr}[
\rho_{m}^{\text{T}_{\text{B}}}
\hat{f}^{\dagger}\hat{f}
]
&
=\frac{1}{4}\sum_{k,
\ell=0}^{1}
\text{Tr}[
\rho_{m,k\ell}^{\text{T}_{\text{B}}}
\hat{f}^{\dagger}\hat{f}]
\notag
\\
&
=\frac{1}{4}\sum_{k,\ell=0}^{1}
\{
z_{i_{1}}^{*}z_{i_{2}}A^{k\ell}_{i_{1}i_{2}}
+
z_{i_{1}}^{*}\zeta_{i_{3}i_{4}}B^{k\ell}_{i_{1},i_{3}i_{4}}
+
z_{i_{3}}\zeta_{i_{1}i_{2}}^{*}(B^{k\ell}_{i_{3},i_{1}i_{2}})^{*}
+
\zeta_{i_{1}i_{2}}^{*}
\zeta_{i_{3}i_{4}}
D^{k\ell}_{i_{1}i_{2},i_{3}i_{4}}
\}
\notag
\\
&
\quad
+\frac{1}{4}\sum_{k,\ell=0}^{1}
\left|
z_{i_1}
(
\langle \hat{r}_{i_1} \rangle^{\text{T}_\text{B}}_{k \ell}
-\langle \hat{r}_{i_1} \rangle^{\text{T}_\text{B}}
)
+
\zeta_{i_1 i_2}
(
\langle \hat{r}_{i_1} \rangle^{\text{T}_\text{B}}_{k \ell}
-\langle \hat{r}_{i_1} \rangle^{\text{T}_\text{B}}
)
(
\langle \hat{r}_{i_2} \rangle^{\text{T}_\text{B}}_{ k \ell}
-\langle \hat{r}_{i_2} \rangle^{\text{T}_\text{B}}
)\right|^{2}\notag\\
&
\ge
\frac{1}{4}\sum_{k,\ell=0}^{1}
\{
z_{i_{1}}^{*}z_{i_{2}}A^{k\ell}_{i_{1}i_{2}}
+
z_{i_{1}}^{*}\zeta_{i_{3}i_{4}}B^{k\ell}_{i_{1},i_{3}i_{4}}
+
z_{i_{3}}\zeta_{i_{1}i_{2}}^{*}(B^{k\ell}_{i_{3},i_{1}i_{2}})^{*}
+
\zeta_{i_{1}i_{2}}^{*}
\zeta_{i_{3}i_{4}}
D^{k\ell}_{i_{1}i_{2},i_{3}i_{4}}
\},
\end{align}
where $\langle 
\hat{r}_i \rangle^{\text{T}_\text{B}}=
\text{Tr}[
\rho_m^{\text{T}_\text{B}} \hat{r}_i
]
$ and 
$\langle 
\hat{r}_i \rangle^{\text{T}_\text{B}}_{k\ell}
=
\text{Tr}[
\rho_{m,k\ell}^{\text{T}_\text{B}}
\hat{r}_i
]$.
The components of the matrix $A^{k\ell}$, $B^{k\ell}$, and $D^{k\ell}$ are
\begin{align}
A^{k\ell}_{i_{1}i_{2}}
&
=\frac{1}{2}\left((\sigma_{kl}^{\text{T}_{\text{B}}})_{i_{1}i_{2}}+i\Omega_{i_{1}i_{2}}\right),
\\
B^{k\ell}_{i_1,i_2 i_3}
&=
A^{k\ell}_{i_{1}i_{2}}
(
\langle \hat{r}_{i_3} \rangle^{\text{T}_\text{B}}_{k \ell}
-\langle \hat{r}_{i_3} \rangle^{\text{T}_\text{B}}
)
+
A^{k\ell}_{i_1 i_3}
(
\langle \hat{r}_{i_2} \rangle^{\text{T}_\text{B}}_{k \ell}
-\langle \hat{r}_{i_2} \rangle^{\text{T}_\text{B}}
)
+
A^{k\ell}_{i_2 i_3}
(
\langle \hat{r}_{i_1} \rangle^{\text{T}_\text{B}}_{k \ell}
-\langle \hat{r}_{i_1} \rangle^{\text{T}_\text{B}}
),
\\
D^{kl}_{i_{1}i_{2},i_{3}i_{4}}
&=
A^{k\ell *}_{i_1 i_2}
(
\langle \hat{r}_{i_3} \rangle^{\text{T}_\text{B}}_{k \ell}
-\langle \hat{r}_{i_3} \rangle^{\text{T}_\text{B}}
)
(
\langle \hat{r}_{i_3} \rangle^{\text{T}_\text{B}}_{k \ell}
-\langle \hat{r}_{i_3} \rangle^{\text{T}_\text{B}}
)
+
A^{k\ell }_{i_1 i_3}
(
\langle \hat{r}_{i_2} \rangle^{\text{T}_\text{B}}_{k \ell}
-\langle \hat{r}_{i_2} \rangle^{\text{T}_\text{B}}
)
(
\langle \hat{r}_{i_4} \rangle^{\text{T}_\text{B}}_{k \ell}
-\langle \hat{r}_{i_4} \rangle^{\text{T}_\text{B}}
)
\nonumber 
\\
&
\quad
+
A^{k\ell }_{i_1 i_4}
(
\langle \hat{r}_{i_2} \rangle^{\text{T}_\text{B}}_{k \ell}
-\langle \hat{r}_{i_2} \rangle^{\text{T}_\text{B}}
)
(
\langle \hat{r}_{i_3} \rangle^{\text{T}_\text{B}}_{k \ell}
-\langle \hat{r}_{i_3} \rangle^{\text{T}_\text{B}}
)
+
A^{k\ell }_{i_2 i_3}
(
\langle \hat{r}_{i_1} \rangle^{\text{T}_\text{B}}_{k \ell}
-\langle \hat{r}_{i_1} \rangle^{\text{T}_\text{B}}
)
(
\langle \hat{r}_{i_4} \rangle^{\text{T}_\text{B}}_{k \ell}
-\langle \hat{r}_{i_4} \rangle^{\text{T}_\text{B}}
)
\nonumber 
\\
&
\quad
+
A^{k\ell }_{i_2 i_4}
(
\langle \hat{r}_{i_1} \rangle^{\text{T}_\text{B}}_{k \ell}
-\langle \hat{r}_{i_1} \rangle^{\text{T}_\text{B}}
)
(
\langle \hat{r}_{i_3} \rangle^{\text{T}_\text{B}}_{k \ell}
-\langle \hat{r}_{i_3} \rangle^{\text{T}_\text{B}}
)
+
A^{k\ell }_{i_3 i_4}
(
\langle \hat{r}_{i_1} \rangle^{\text{T}_\text{B}}_{k \ell}
-\langle \hat{r}_{i_1} \rangle^{\text{T}_\text{B}}
)
(
\langle \hat{r}_{i_2} \rangle^{\text{T}_\text{B}}_{k \ell}
-\langle \hat{r}_{i_2} \rangle^{\text{T}_\text{B}}
)
\nonumber 
\\
&
\quad
+
A^{k\ell *}_{i_1 i_2}
A^{k\ell }_{i_3 i_4}
+
A^{k\ell }_{i_1 i_3}
A^{k\ell }_{i_2 i_4}
+
A^{k\ell }_{i_1 i_4}
A^{k\ell }_{i_2 i_3}, 
\end{align}
where 
$(
\sigma^{\text{T}_\text{B}}_{kl}
)_{ij} 
=
\text{Tr}[
\rho_{m,kl}^{\text{T}_{\text{B}}}
\{
\hat{r}_i -
\langle 
\hat{r}_i \rangle^{\text{T}_\text{B}}_{k\ell},
\hat{r}_j -
\langle 
\hat{r}_j \rangle^{\text{T}_\text{B}}_{k\ell}
\}
]$ is the covariance matrix of each Gaussian state, which does not depend on the coupling $\Lambda$.
Moreover, the difference
$\langle 
\hat{r}_i \rangle^{\text{T}_\text{B}}_{k\ell}
-\langle 
\hat{r}_i \rangle^{\text{T}_\text{B}}$
between the expectation values 
is proportional to $\Lambda$. 
Hence, the right-hand side of the inequality \eqref{bound} is up to the order of $\Lambda^{2}$.
Fig. \ref{fig:LL} shows the behavior of the minimum eigenvalue as a function of $\Lambda$. This demonstrates that the minimum eigenvalue is proportional to $\Lambda^{2}$,
which follows the lower bound estimated by the inequality \eqref{bound}.

\section{SUMMARY AND CONCLUSION}
We developed an entanglement criterion based on the covariance matrix and third- and fourth-order cumulants.
The higher-order cumulants characterize the non-Gaussianity; hence, this condition enables us to evaluate the entanglement even for non-Gaussian states.
We applied the formula to a cat-like state, which manifests as a non-Gaussian entangled state.
We showed that the criterion using up to the fourth-order cumulant is useful for characterizing the entanglement of the system with non-Gaussianity.
 
We applied the formula to an optomechanical system to investigate 
the gravity-induced entanglement between mechanical mirrors. 
We demonstrated how the entanglement between the two mirrors coupled 
to cavity photons is detected by the formula. 
The quantum superposition of the position of the mirrors caused by the interaction with the cavity photons leads to non-Gaussian features.
The cumulant-based condition facilitates the detection of the entanglement between the mirrors in the non-Gaussian state.
Furthermore, we discussed the crucial role of third- and fourth-order cumulants to observe the entanglement.
However, for the highly non-Gaussian state, the method to detect the entanglement is limited because cumulants higher than the fourth-order one become more important to describe such a non-Gaussian state.

To apply this cumulant-based condition to an experimental setup, 
we need to measure the correlation between the position and the momentum up to the 
fourth order. 
Several measurements of the correlations have been discussed experimentally.
For example, Ref.~\cite{Matsumoto2} measured the correlation between the position 
and momentum of a single suspended mirror interacting with photons.
Ref.~\cite{Clarke} proposed a measurement scheme to measure the second-order moments 
of two mechanical oscillators.
By extending these measurements to the fourth-order moments, in principle, our scheme 
for the entanglement criterion can be applied to detect gravity-induced entanglement. 
However, the feasibility of entanglement detection will be studied in the future.

\appendix
\section{\label{app:level1}RELATION BETWEEN CUMULANT AND MOMENT}
In this appendix, we derive the cumulants from the characteristic function for any state $\rho$.
Here, we consider the series expansion of the natural logarithm of the characteristic function
\begin{align}
\label{chaex}
\text{ln}\chi(\bm{r})
&=ir_{j}\Omega_{jk}\kappa_{1,k}+\frac{i^2}{2}r_{j}\Omega_{jk}r_{l}\Omega_{lm}\kappa_{2,lm}+\cdots,
\end{align}
where the coefficients $\kappa_{n}$ are known as the $n$th-order cumulants.
Then, we obtain the $n$th-order cumulant from the $n$th-order derivative of Eq. \eqref{chaex} at a point $\bm{r}={\bm  0}$.
Using the characteristic function $\chi(\bm{r})=\text{Tr}[\rho e^{i\bm{r}^{T}\Omega\hat{\bm{r}}}]$, we derive the first- to fourth-order cumulants as
\begin{align}
\begin{aligned}
&\kappa_{1,i}=\text{Tr}[\rho\hat{r}_{i}]\\
&\kappa_{2,ij}=\frac{1}{2}\text{Tr}[\rho\{\Delta\hat{r}_{i},\Delta\hat{r}_{j}\}]\\
&\kappa_{3,ijk}=\text{Tr}[\rho\Delta\hat{r}_{(i}\Delta\hat{r}_{j}\Delta\hat{r}_{k)}]\\
&\kappa_{4,ijkl}
=\text{Tr}[\rho\Delta\hat{r}_{(i}\Delta\hat{r}_{j}\Delta\hat{r}_{k}\Delta\hat{r}_{l)}]-(\kappa_{2,ij}\kappa_{2,kl}+\kappa_{2,ik}\kappa_{2,jl}+\kappa_{2,il}\kappa_{2,jk}),
\end{aligned}
\end{align}
where $\Delta\hat{r}_{i}=\hat{r}_{i}-\text{Tr}[\rho\hat{r}_{i}]$ and the subscript brackets represent the symmetry, such as
\begin{align}
\label{sym}
\hat{r}_{(i}\hat{r}_{j}\hat{r}_{k)}
&=\frac{\hat{r}_{i}\hat{r}_{j}\hat{r}_{k}+\hat{r}_{i}\hat{r}_{k}\hat{r}_{j}
+\hat{r}_{j}\hat{r}_{i}\hat{r}_{k}+\hat{r}_{j}\hat{r}_{k}\hat{r}_{i}
+\hat{r}_{k}\hat{r}_{i}\hat{r}_{j}+\hat{r}_{k}\hat{r}_{j}\hat{r}_{i}}{3!}.
\end{align}
Furthermore, the first- to fourth-order central moments are given by
\begin{align}
\begin{aligned}
&m_{1,i}=\braket{\hat{r}_{i}}
=\text{Tr}[\rho\hat{r}_{i}],\\
&m_{2,ij}=\sigma_{ij}
=\text{Tr}[\rho\{\Delta\hat{r}_{i},\Delta\hat{r}_{j}\}]\\
&m_{3,ijk}=\text{Tr}[\rho\Delta\hat{r}_{(i}\Delta\hat{r}_{j}\Delta\hat{r}_{k)}]\\
&m_{4,ijkl}
=\text{Tr}[\rho\Delta\hat{r}_{(i}\Delta\hat{r}_{j}\Delta\hat{r}_{k}\Delta\hat{r}_{l)}],
\end{aligned}
\end{align}
and the cumulants are expressed by the central moments as
\begin{align}
\begin{aligned}
&\kappa_{1,i}=m_{1,i}\\
&\kappa_{2,ij}=\frac{1}{2}m_{2,ij}\\
&\kappa_{3,ijk}=m_{3,ijk}\\
&\kappa_{4,ijkl}
=m_{4,ijkl}-(m_{2,ij}m_{2,kl}+m_{2,ik}m_{2,jl}+m_{2,il}m_{2,jk}).
\end{aligned}
\end{align}

\section{\label{app:level2}OPERATION OF PARTIAL TRANSPOSITION}
In this section, we describe the partial transposition of the cumulants.
In the Fock basis, the transposition of the canonical operators $\hat{q}$ and $\hat{p}$ is given by
\begin{align}
\hat{q}^{\text{T}}
&=\frac{(\hat{a}_{1}+\hat{a}_{1}^{\dagger})^{\text{T}}}{\sqrt{2}}=\hat{q},\quad
\hat{p}^{\text{T}}
=\frac{(\hat{a}_{1}-\hat{a}_{1}^{\dagger})^{\text{T}}}{i\sqrt{2}}=-\hat{p},
\end{align}
where $\hat{a}_{1}$ and $\hat{a}_{1}^{\dagger}$ are the creation and annihilation operators, respectively.
Thus, the partial transposition of the vector of the canonical operators $\hat{\bm{r}}=(\hat{q}_{a},\hat{p}_{a},\hat{q}_{b},\hat{p}_{b})^{\text{T}}$ is
\begin{align}
\hat{\bm{r}}^{\text{T}_{\text{B}}}
&=T\hat{\bm{r}},\quad
T=\text{diag}(1,1,1,-1).
\end{align}

Here, the Weyl ordered operator $O(\hat{q}_{a},\hat{p}_{a},\hat{q}_{b},\hat{p}_{b})$, which is the average of all possible products of the position and momentum operators, satisfies
\begin{align}
\label{normal}
\text{Tr}[\rho O(\hat{q}_{a},\hat{p}_{a},\hat{q}_{b},\hat{p}_{b})]
&=\int O(q_{a},p_{a},q_{b},p_{b})W(q_{a},p_{a},q_{b},p_{b})d^{2}\bm{r},
\end{align}
where $W(q_{a},p_{a},q_{b},p_{b})$ is the Wigner function.
Under the partial transposition with respect to system B, the Wigner function changes to $W(q_{a},p_{a},q_{b},p_{b})\rightarrow W(q_{a},p_{a},q_{b},-p_{b})$.
Hence, we rewrite the relation \eqref{normal} for the partial transposed density matrix as
\begin{align}
\text{Tr}[\rho^{\text{T}_{\text{B}}}O(\hat{q}_{a},\hat{p}_{a},\hat{q}_{b},\hat{p}_{b})]
&=\int O(q_{a},p_{a},q_{b},p_{b})W(q_{a},p_{a},q_{b},-p_{b})d^{2}\bm{r},\notag\\
&=\text{Tr}[\rho O(\hat{q}_{a},\hat{p}_{a},\hat{q}_{b},-\hat{p}_{b})].
\end{align}
From this relation, we derive the partial transposition of the cumulants as
\begin{align}
\begin{aligned}
\kappa_{2,ij}^{\text{T}_{\text{B}}}
&=T_{ii'}T_{jj'}\kappa_{2,i'j'}\\
\kappa_{3,ijk}^{\text{T}_{\text{B}}}
&=T_{ii'}T_{jj'}T_{kk'}\kappa_{3,i'j'k'}\\
\kappa_{4,ijkl}^{\text{T}_{\text{B}}}
&=T_{ii'}T_{jj'}T_{kk'}T_{ll'}\kappa_{4,ijkl}.
\end{aligned}
\end{align}

\section{\label{app:level3}DERIVATION OF CUMULANT FOR OPTOMECHANICAL SYSTEM}
In this appendix, we derive the first- to fourth-order cumulants for the optomechanical system.
The $n$-th-order cumulant is defined by Eq. \eqref{ncum} and the characteristic function 
of the optomechanical system is obtained by Eq. \eqref{cha}.
Using these equations, we derive the cumulants as
\begin{align}
\kappa_{1,i}
&=-\frac{1}{4}\sum_{n,m=0}^{1}S_{ij}(j_{nm,j}'+r'_{j})
=\frac{1}{4}\sum_{n,m=0}^{1}\braket{r_{i}}_{nm},\\
\kappa_{2,ij}
&=\frac{1}{2}\left(S_{ik}S^{T}_{kj}
+\frac{1}{2}\sum_{n,m=0}^{1}\braket{\hat{r}_{i}}_{nm}\braket{\hat{r}_{j}}_{nm}
-2\braket{\hat{r}_{i}}\braket{\hat{r}_{j}}\right),\\
\kappa_{3,ijk}
&=-\frac{1}{4}\sum_{n,m=0}^{1}\eta_{nm,i}\eta_{nm,j}\eta_{nm,k}-\frac{2}{64}\sum_{n,m=0}^{1}\sum_{n',m'=0}^{1}\sum_{n'',m''=0}^{1}\eta_{nm,i}\eta_{n'm',j}\eta_{n''m'',k}\notag\\
&\quad+\frac{1}{16}\sum_{n,m=0}^{1}\sum_{n',m'=0}^{1}\left(\eta_{nm,i}\eta_{nm,j}\eta_{n'm',k}+\eta_{nm,i}\eta_{nm,k}\eta_{n'm',j}+\eta_{nm,j}\eta_{nm,k}\eta_{n'm',i}\right),
\end{align}
\begin{align}
\label{4cum}
\kappa_{4,ijkl}
&=\frac{1}{4}\sum_{n,m=0}^{1}\eta_{nm,i}\eta_{nm,j}\eta_{nm,k}\eta_{nm,l}
-\frac{6}{256}\sum_{n,m=0}^{1}\sum_{n',m'=0}^{1}\sum_{n'',m''=0}^{1}\sum_{n''',m'''=0}^{1}
\eta_{nm,i}\eta_{n'm',j}\eta_{n''m'',k}\eta_{n'''m''',l}\notag\\
&\quad-\frac{1}{16}\sum_{n,m=0}^{1}\sum_{n',m'=0}^{1}(
\eta_{nm,i}\eta_{nm,j}\eta_{n'm',k}\eta_{n'm',l}
+\eta_{nm,i}\eta_{n'm',j}\eta_{nm,k}\eta_{n'm',l}
+\eta_{nm,i}\eta_{n'm',j}\eta_{n'm',k}\eta_{nm,l})\notag\\
&\quad-\frac{1}{16}\sum_{n,m=0}^{1}\sum_{n',m'=0}^{1}(
\eta_{nm,i}\eta_{nm,j}\eta_{nm,k}\eta_{n'm',l}
+\eta_{nm,i}\eta_{nm,j}\eta_{n'm',k}\eta_{nm,l}\notag\\
&\quad\quad\quad\quad\quad\quad\quad
+\eta_{nm,i}\eta_{n'm',j}\eta_{nm,k}\eta_{nm,l}
+\eta_{n'm',i}\eta_{nm,j}\eta_{nm,k}\eta_{nm,l})\notag\\
&\quad+\frac{2}{64}\sum_{n,m=0}^{1}\sum_{n',m'=0}^{1}\sum_{n'',m''=0}^{1}(
\eta_{nm,i}\eta_{nm,j}\eta_{n'm',k}\eta_{n''m'',l}
+\eta_{nm,i}\eta_{n'm',j}\eta_{nm,k}\eta_{n''m'',l}\notag\\
&\quad\quad\quad\quad\quad\quad\quad\quad\quad\quad
+\eta_{nm,i}\eta_{n'm',j}\eta_{n''m'',k}\eta_{nm,l}
+\eta_{n'm',i}\eta_{nm,j}\eta_{nm,k}\eta_{n''m'',l}\notag\\
&\quad\quad\quad\quad\quad\quad\quad\quad\quad\quad
+\eta_{n'm',i}\eta_{nm,j}\eta_{n''m'',k}\eta_{nm,l}
+\eta_{n'm',i}\eta_{n''m'',j}\eta_{nm,k}\eta_{n''m'',l}\Big),
\end{align}
where $\bm{j}'_{kl}=H^{-1}(1-e^{-tH\Omega})\bm{j}_{kl}$ and $\bm{j}_{kl}=(k\Lambda\Omega_{m}(1+g)^{-\frac{1}{4}},0,l\Lambda\Omega_{m}(1+g)^{-\frac{1}{4}},0)^{\text{T}}$.
Here, we define $\bm{\eta}_{nm}=S\bm{j}_{nm}'$ and the first-order moment of each Gaussian state  $\braket{\hat{\bm{r}}}_{kl}=S(\bm{j}_{kl}'+\bm{r}')$.

\end{document}